\documentclass[12pt]{article}

\input epsf
\usepackage{amssymb,amsmath}
\usepackage{wrapfig,subfigure}
\usepackage[matrix,arrow,curve]{xy}
\usepackage{cite}
\usepackage{pdfsync}
\usepackage{graphicx,color}
\usepackage[debug,pageanchor=false]{hyperref}
\definecolor{link}{rgb}{.8,.15,.1}
\hypersetup{colorlinks=true,linkcolor=link,citecolor=link,urlcolor=link,linktocpage}
\usepackage{cancel}

\makeatletter
\@addtoreset{equation}{section}
\makeatother

\def\cc {{\mathbb{C}}}
\def\rr {{\mathbb{R}}}
\def\zz {{\mathbb{Z}}}
\def\del {\partial}

\def\del {\partial}

\def\lsim{\mathrel{\rlap{\lower4pt\hbox{\hskip1pt$\sim$}}
    \raise1pt\hbox{$<$}}}                

\begin{document}

\begin{titlepage}

\begin{center}

\vskip .3in \noindent

{\Large \bf{Holography for $(1,0)$ theories in six dimensions}}

\bigskip

	Davide Gaiotto$^1$ and Alessandro Tomasiello$^2$\\

       \bigskip
	{$^1$ Perimeter Institute for Theoretical Physics, Waterloo, Ontario, Canada N2L 2Y5 \\
	\vspace{.1cm} 
	$^2$ Dipartimento di Fisica, Universit\`a di Milano--Bicocca, I-20126 Milano, Italy\\
    and\\
    INFN, sezione di Milano--Bicocca
	}

       \vskip .5in
       {\bf Abstract }
       \vskip .1in
\end{center}
M-theory and string theory predict the existence of many six-dimensional SCFTs. In particular, type IIA brane constructions involving NS5-, D6- and D8-branes 
conjecturally give rise to a very large class of ${\cal N}=(1,0)$ CFTs in six dimensions. We point out that these theories sit at the end of RG flows which start from 
six-dimensional theories which admit an M-theory construction as a M5 stack transverse to $\rr^4/\zz_k \times \rr$. 
The flows are triggered by Higgs branch expectation values and correspond to D6's opening up into transverse D8-branes via a Nahm pole. 
We find a precise correspondence between these CFT's and the AdS$_7$ vacua found in a recent classification in type II theories. 
Such vacua involve massive IIA regions, and the internal manifold is topologically $S^3$. 
They are characterized by fluxes for the NS three-form and RR two-form, which can be thought of as the near-horizon version of the NS5's and D6's in the brane picture; 
the D8's, on the other hand, are still present in the AdS$_7$ solution, in the form of an arbitrary number of concentric shells wrapping round $S^2$'s.  

\noindent

\vfill
\eject

\end{titlepage}

\hypersetup{pageanchor=true}

\tableofcontents

\section{Introduction} 
\label{sec:intro}

There are various reasons to be interested in six-dimensional conformal field theories (CFTs): the intrinsic field-theoretic interest of defining higher-dimensional theories, the fact that one such theory lives on the M5 worldvolume, and the great variety of interesting field theories they give rise to upon compactification \cite{gaiotto}. 

The $(2,0)$ theories are conjecturally classified by a choice of an ADE ``gauge group''. Some $(1,0)$ theories can be understood as ``orbifolds'' of $(2,0)$ theories. 
More interesting possibilities were pointed out in \cite{hanany-zaffaroni-6d,brunner-karch}: their theories consisted of intersecting D8--D6--NS5 systems, 
similar to the ones considered to great effect in \cite{hanany-witten,witten-D4} to study three-dimensional and four-dimensional theories. 
It was possible for their NS5's to coincide, which suggested the appearance of tensionless strings and hence the emergence of CFTs. 

It would clearly be interesting to have a classification of $(1,0)$ theories. Recently, this problem was attacked in \cite{afrt} by classifying AdS$_7 \times M_3$ vacua in type II supergravity. While in IIB no solution was found,\footnote{In F-theory, which goes beyond IIB supergravity, \cite{heckman-morrison-vafa} recently found many $(1,0)$ theories, and classified them. One of their series admitting a large $N$ limit consists of the $(2,0)$ theory $A_N$ connected to two D7 ``tails''; this is formally reminiscent of our construction in this paper. It would be interesting to understand the relationship more precisely.} many new solutions emerged in massive IIA.\footnote{As we will see, for all these solutions the internal space is an $S^2$-fibration over an interval. This Ansatz had been considered earlier: \cite{blaback-danielsson-junghans-vanriet-wrase-zagermann,gautason-junghans-zagermann} considered smeared solutions and pointed out problems with source localization (for an updated discussion of which see section 5.2 of \cite{afrt}); \cite{blaback-danielsson-junghans-vanriet-wrase-zagermann-2} found local solutions. More recently, \cite{junghans-schmidt-zagermann} studied the existence of solutions with a single D8 stack, such as the one in figure \ref{fig:1d8}, from the point of view of the brane effective potential, and suggested the existence of non-supersymmetric solutions as well.} This at least gives a classification of theories having a IIA dual. 

The massive solutions of \cite{afrt} have an internal manifold homeomorphic to $S^3$; the active form fields are the NS three-form $H$, and the RR fields $F_0$, $F_2$.  In general they involve several concentric D8 shells (with non-zero D6-charge). It is natural to conjecture that these solutions are somehow the near-horizon limits of the D8--D6--NS systems used in \cite{hanany-zaffaroni-6d,brunner-karch}. This conjecture cannot be checked directly, since the solution for such a brane intersection has never been found (although some steps in that direction were made for example in \cite{janssen-meessen-ortin,imamura-D8,youm}). However, in this paper we provide evidence for it by studying the correspondence in some detail. 

We first point out that the CFTs arise by going on special loci of the Higgs moduli space of the $A_N$ (1,0) ``orbifold'' theory. We use methods initially developed for the study of BPS boundary conditions in ${\cal N}=4$ super-Yang--Mills (SYM) \cite{gaiotto-witten-1,gaiotto-witten-3}. We first start with a stack of coinciding NS5's on a stack of D6's, and then decouple the seven-dimensional degrees of freedom of the latter by having the D6's end on two stacks of D8's realizing Dirichlet boundary conditions. We can then add a Higgs deformation that changes the brane configuration to a more general one, and that lets us flow to a more general CFT. 

We then refine the classification of AdS$_7$ solutions in \cite{afrt}, to check that there is a one-to-one correspondence with the brane-intersection CFT's. In \cite{afrt} a general characterization was found, along with some examples. Here we find evidence that infinitely many solutions exist, and we give an empirical classification that happens to be in one-to-one correspondence with the CFT's coming from the D8--D6--NS5 system. For example, we find that there is a bound on the flux integer $N$ of the NS three-form $H$, corresponding to a bound on the number of NS5 branes on the CFT side. Moreover, the D8's are ordered in such a way that the ones with lower D6 charge are pushed towards the poles; this reproduces a constraint on the brane ordering first found in the context of D5--D3--NS5 systems in \cite{gaiotto-witten-1}. 

We also use the AdS duals to give an estimate of the number of degrees of freedom of our CFT's. This cannot be easily done in general, because our solutions are numerical. However, we can perform the computation in certain limits. For example, the number of degrees of freedom of the ``massless'' $A_k$ theory goes like $N^3 k^2$; for the easiest type of solutions, with Romans mass quantized by $F_0 = 2 \pi n_0$, we show that the first correction to this goes like $N \frac{k^4}{n_0^2}$. 

The paper is organized as follows. In section \ref{sec:cft} we discuss the six-dimensional CFTs, both from an abstract viewpoint and from their origin as D8--D6--NS5 systems. We then consider the gravity side: in section \ref{sec:ads} we review the results of \cite{afrt} about AdS$_7$ vacua in massive IIA, and in section \ref{sec:expl} we refine that classification to show evidence for a bijective correspondence with the CFTs discussed in section \ref{sec:cft}. Finally, in section \ref{sec:free}, we use the AdS solutions to compute the number of degrees of freedom for our theories in some simple cases. 
 

\section{On $(1,0)$ 6d SCFTs } 
\label{sec:cft}

There is a variety of $(1,0)$ 6d SCFTs which arise from a low energy limit of the world-volume theory of M5 branes in the presence of other half-BPS objects
in M-theory. These theories are related by an intricate network of RG flows, initiated by turning on vevs in either of the two branches of vacua of these theories, 
the Higgs branch and the tensor branch. 
Many such flows have a simple interpretation in the M-theory construction as separating the M5 branes from each other or from the other half-BPS objects. 
In this section we would like to describe some flows which do not admit such an interpretation, and 
seem hard to realize directly in an M-theory UV description of the system. 
At the endpoint of the flows we will find the $(1,0)$ 6d SCFTs whose existence 
was predicted by Hanany and Zaffaroni \cite{hanany-zaffaroni-6d}, based on IIA brane-engineering constructions. 

Along the way we will use some techniques (notably the description of intersecting branes as Nahm poles) which were discussed extensively in the context of lower-dimensional field theories in \cite{gaiotto-witten-1,gaiotto-witten-3}, to which we refer the reader for background. 

\subsection{SCFTs from brane systems and Higgs branch RG flows} 
\label{sub:scft}

The most basic M-theory setup involves $N$ M5 branes sitting at the locus of an A-type singularity $\rr^4/\zz_k$. 
If the singularity wraps the $x^0$ to $x^6$ directions, the M5 wrap the $x^0$ to $x^5$ directions.
The singularity itself supports a seven-dimensional 
${\rm SU}(k)$ gauge theory and the M5 branes behave as domain walls in the world-volume of the gauge theory. 
Thus the resulting six-dimensional SCFTs, which we will call here $T^{A_N}_{A_k}$, are characterized by a ${\rm SU}(k)_L \times {\rm SU}(k)_R$ flavor symmetry which is coupled to the 7d gauge fields 
in the M-theory UV setup. The coupling to the 7d gauge theory implicit in the M-theory setup is part of the reason not all Higgs branch deformations of the six-dimensional SCFT 
are visible in M-theory: the Higgs branch of the combined 6d--7d system is roughly an ${\rm SU}(k) \times {\rm SU}(k)$ hyper-K\"ahler quotient of the 6d Higgs branch.

\begin{figure}[ht]
\centering	
\includegraphics[scale=.4]{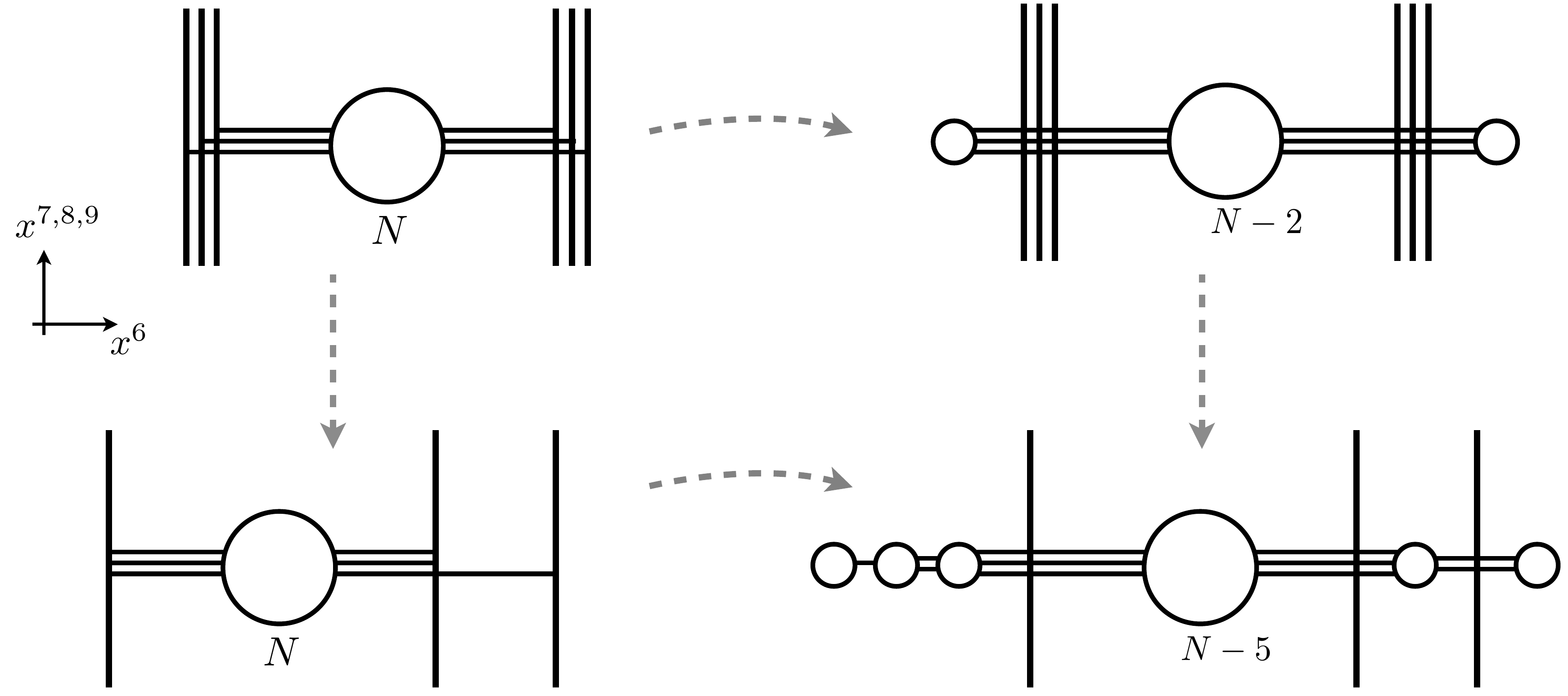}
	\caption{Brane configurations realizing some of the theories we consider in this paper. The top left configuration represents the theory $T^{A_N}_{A_k}$, describing the interaction of $N$ NS5-branes with $k$ D6's; the D8-branes on the two sides enforce Dirichlet boundary conditions for the fields on the D6's, thus decoupling the seven-dimensional degrees of freedom from the six-dimensional ones. The other brane configurations are induced by Higgs branch deformations (vertical arrows) or tensor branch deformations (horizontal arrows).}
	\label{fig:branes}
\end{figure}

In order to decouple the seven-dimensional gauge theory, we can move from an M-theory to a IIA string theory description of the system, 
simply by replacing $\rr^4/\zz_k$ with a charge $k$ Taub--NUT geometry. Upon reducing to IIA, the Taub--NUT geometry gives a bundle of $k$ D6 branes and the M5 branes reduce to $N$ NS5 branes sitting on the D6 branes. 
The six-dimensional SCFT is now essentially described as the world-volume theory on the NS5 branes. The advantage of the IIA description is that one can modify the setup to add Dirichlet boundary conditions for the seven-dimensional gauge theory on the D6 branes, by having them end on two stacks of $k$ D8 branes places at large positive and large negative $x^6$.
More precisely, each D6 will end on a separate D8 brane, so that if we separate the D8 branes slightly in the $x^6$ direction, 
the number of D6 brane segments in each interval decreases linearly from $k$ to $0$ as one moves across the D8 brane stack; see the top left configuration in figure \ref{fig:branes}. 
Notice that this construction requires massive IIA with $F_0 = \pm k$ for very large positive and negative $x^6$. We do not know how to mimic that in M-theory. 
The construction is a six-dimensional version of the brane systems used in \cite{gaiotto-witten-1} to describe boundary conditions for ${\cal N}=4$ SYM.

The addition of the D8 branes decouples the six-dimensional degrees of freedom from the seven-dimensional gauge theory, and makes new Higgs branch deformations of the 6d theory visible as the motion of D6 brane segments stretched between D8 branes.

As in the four-dimensional case, the RG flows initiated by such Higgs branch deformations correspond to removing certain subsets of D6-brane segments
by bringing them to infinity along the D8 branes. The result is precisely the general D8--D6--NS5 brane system used by \cite{hanany-zaffaroni-6d} to predict the existence of new 6d SCFTs. What we gain here is a more precise field-theoretic understanding of these constructions. 

In the language of the seven-dimensional gauge theory, the result of the Higgsing procedure is to replace the Dirichlet boundary conditions with a Nahm pole variant, where the three scalar fields $X^i$ of the gauge theory blow up as 
\begin{equation}\label{eq:nahm}
X^i \sim \pm \frac{t^i_\rho}{x^6 - x^6_\partial}
\end{equation}
at the boundary, with $t^i_\rho$ being the generators of an $su(2)$ subgroup of $SU(k)$ labelled by the 
partition (or better, $su(2)$ embedding) $\rho$. 

As in \cite{gaiotto-witten-1,gaiotto-witten-3}, the correspondence between the brane system and the Nahm pole picture works most easily if the D8-branes obey an ``ordering constraint''. Namely, the number of D6 ending on each D8 is a decreasing function as we get further away from the central NS5 stack. (For example it is obeyed in the lower-left picture in figure \ref{fig:branes}; on the right of the central stack, the number of D6's ending on the D8's are 2 and 1.) We will see in section \ref{ssub:order} that this emerges very clearly from the gravity duals.

As seven-dimensional gauge theory is not UV complete, a pure field-theoretic description of the Higgsing procedure should be formulated in terms of the original 6d SCFT and its Higgs branch. In analogy with the four-dimensional construction of three-dimensional SCFTs, we expect the relevant Higgs branch deformations of the six-dimensional SCFT to be parameterized by the complex moment maps $\mu^{\mathbb{C}}_{L,R}$ for the ${\rm SU}(k)_L \times {\rm SU}(k)_R$ flavor symmetry of the SCFT. The effect of the Nahm pole in the 7d gauge theory is to force the moment map vevs to lie in a non-generic nilpotent orbit $\bar O_\rho$
labelled by the corresponding su(2) subgroup. 

At the bottom of the RG flow, we may hope to find a six-dimensional SCFT labelled by the two embeddings, $\rho_{L,R}$, chosen at the two boundaries for the seven-dimensional system; we will denote them as $T^{A_N}_{A_k,\rho_L, \rho_R}$. The most obvious criterion to understand the behavior of the RG flow is to look at the geometry of the Higgs branch in the neighborhood of the direction we deform along: if the normal geometry is sufficiently singular, the low energy theory will have a non-trivial Higgs branch geometry and should thus be an SCFT. 
If some normal directions are non-singular, they should be parameterized by free hypermultiplets in the IR. If the normal geometry 
is non-singular, the low energy theory likely consists of free fields only. 

The Higgs branch geometry can be computed from the gauge theory description of the brane system exactly as in the lower dimensional case \cite{gaiotto-witten-1} and thus the condition on the 
choices of $\rho_{L,R}$, $k$ and $N$ to be compatible with a smooth flow to a new 6d SCFT $T^{A_N}_{A_k,\rho_L, \rho_R}$ are likely the same as in the lower dimensional case as well. 
We will discuss them momentarily.


\subsection{Quivers on the tensor branch} 
\label{sub:branes}

The D8--D6--NS5 system \cite{hanany-zaffaroni-6d} is in many ways analogous to the D6--D4--NS5 system used to engineer four-dimensional ${\cal N}=2$ quiver gauge theories \cite{witten-D4}
or the D5--D3--NS5 system used to engineer three-dimensional ${\cal N}=4$ quiver gauge theories \cite{hanany-witten}.
Indeed, if we separate the NS5 branes completely along the $x^6$ direction, the low energy description of the system involves precisely the same linear quiver gauge theories, but in six dimensions. 
An important difference is that the separation between the NS5 branes here survives as a dynamical (tensor multiplet) field rather than 
becoming a coupling. 

The $x^6$ position of D8 branes does not affect the low-energy theory, as long as we take into account the Hahany--Witten effect whenever a D8 brane is carried across an NS5 brane, creating or annihilating the appropriate number of D6 brane segments between the D8 and the NS5 brane. If the D8 branes are brought appropriately to the left or to the right of 
the NS5 branes, we can read off the $\rho_{L,R}$ labels for the $T^{A_N}_{A_k,\rho_L,\rho_R}$ which emerges when all branes are coincident. 

If we separate the NS5 branes fully and move the D8 branes to a configuration with no D6 branes segments ending on D8 branes, but ending on NS5 branes only,
we reach a quiver gauge theory description. Each bundle of $n_a$ D6 brane segments give an ${\rm SU}(n_a)$ gauge group and each D8 brane in the $a$-th interval adds a flavor to the $a$-th gauge group. 
Each individual NS5 brane gives a bi-fundamental hypermultiplet connecting the two nearby gauge groups. 
Each 6d gauge coupling must be promoted to the scalar field in a tensor  multiplet which describes the relative separation of the NS5 branes. 
This is crucial for anomaly cancellation.  The resulting 6d gauge theory is the low energy description of $T^{A_N}_{A_k,\rho_L,\rho_R}$ on its tensor
branch of vacua. 

If the NS5 branes are separated into groups of coincident branes, possibly in the presence of D8 branes inserted in the group, we 
arrive at intermediate effective descriptions of the theory on a sub-branch of the full tensor branch. The descriptions consist of six-dimensional gauge groups coupled to strongly-coupled 6d matter theories. Indeed, a group of coincident branes gives a domain wall between consecutive pairs SU$(n_a)$ and SU$(n_{a+1})$ of gauge groups. Each domain wall is described by coupling the SU$(n_a)$ and SU$(n_{a+1})$ gauge fields to a certain 6d matter theory, a simpler representative in the class of theories considered here. 

\begin{figure}[ht]
\centering	
\includegraphics[scale=.4]{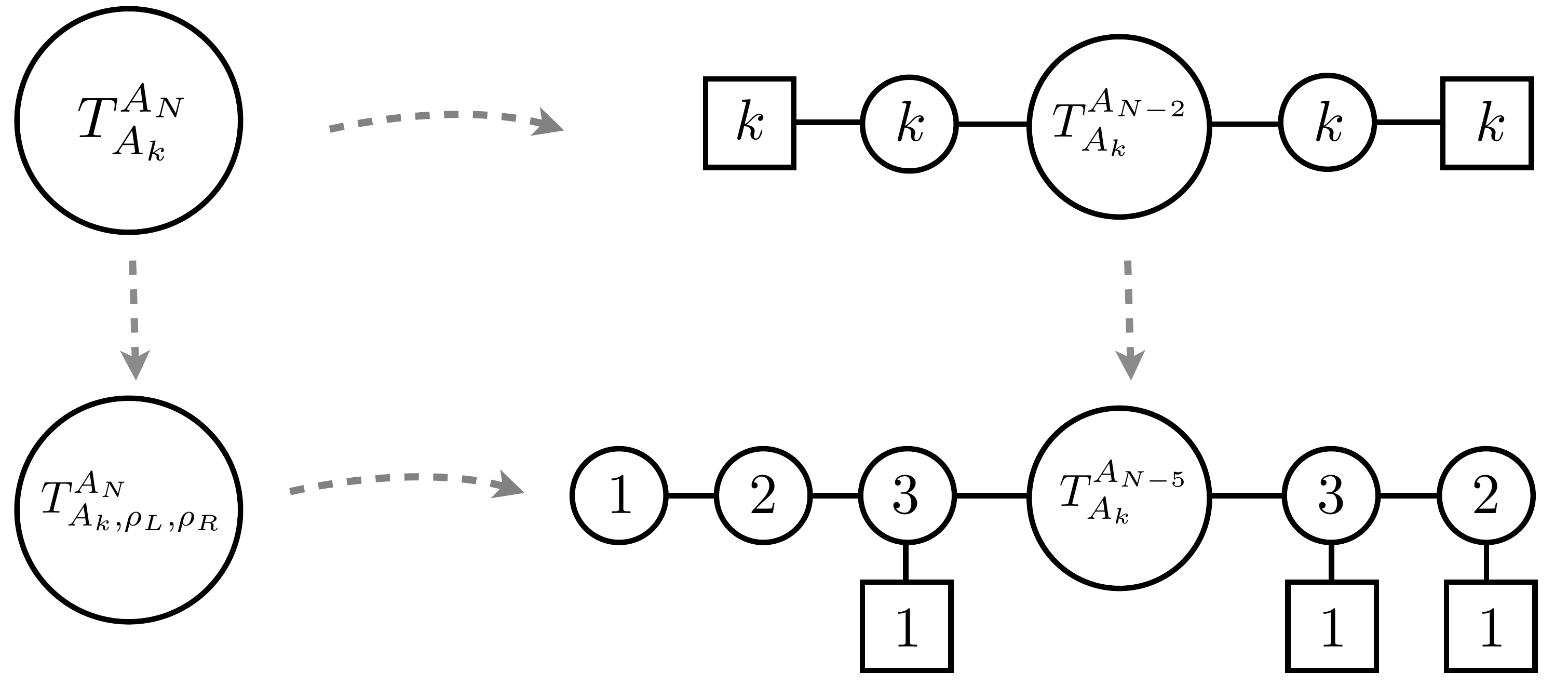}
	\caption{Diagrams describing the field theories corresponding to the brane configurations in figure \ref{fig:branes}. (In that figure there are $k=3$ D6's. The lower-right quiver is appropriate for $k=3$ and the particular $\rho$, $\rho'$ corresponding to the lower-right configuration in figure \ref{fig:branes}.) In each quiver, a small node corresponds to a vector multiplet; a link between small nodes to a bi-fundamental hypermultiplet; 
	a bigger node between two small nodes describes a non-abelian tensor theory $T$ whose flavor symmetries are gauged at the small nodes. The gauge coupling at each small node is promoted to a tensor multiplet scalar. Again the vertical arrows represent RG flows induced by Higgs branch deformations, and the horizontal ones by tensor branch deformations.}
	\label{fig:quivers}
\end{figure}

We can summarize graphically the field theory corresponding to a brane configuration as in figure \ref{fig:quivers}. As in other dimensions, a node describes a vector multiplet, with a number that describes its rank\footnote{The quivers in the figure describe gauge theories with groups U$(n_a)$; the U$(1)$'s in each of them, however, decouples in the IR, as described in \cite{hanany-zaffaroni-6d}.}; a square is used to denote a flavor group. A link between two small nodes describes a bi-fundamental hypermultiplet. We also include bigger nodes representing 6d SCFTs coupled to the gauge fields.
A link between a big node an a small node represents such a coupling. 

The anomaly-cancellation mechanism in the six-dimensional gauge theory also requires $N_f = 2 N_c$ at each node. In the brane setup, this anomaly cancellation condition originates 
from the observation that the net number of D6 branes ending on an NS5 brane must coincide with the value of $2\pi F_0$ 
in the region where the NS5 brane lies:
\begin{equation}\label{eq:netD6}
	n_{\rm D6,\, left}-n_{\rm D6,\, right} = n_0 \equiv 2\pi F_0\ .
\end{equation}
This constraint comes from the Bianchi identity $d F_2 - H F_0 = \delta_{\rm D6}$, integrated around the D6's before and after an NS5. (Recall that flux quantization implies $F_0 =\frac{n_0}{2\pi}$, $n_0 \in \zz$). One can check that all the brane configurations in \ref{fig:branes} satisfy (\ref{eq:netD6}); recall that the region with $F_0=0$ is the one where the large NS5 stack is sitting. 

Let us now denote the numbers of flavors as $m_a$. As $n_{a+1} - n_a$ is the net number of D6 branes ending on the $a$-th NS5 brane, i.e. the value of $2\pi F_0$ there, and a D8 brane sources a unit of $F_0$, we find the expected $N_f = 2 N_c$ relation at each node: $n_{a+1} - 2 n_a + n_{a-1} + m_a = 0$. 

It is convenient to associate to each linear quiver which satisfies these constraints a convex ``Newton polygon'', with vertices at $(a, n_a)$. (See for instance \cite[App.~E]{gaiotto-moore-neitzke} for a quick introduction and some examples.) Then the slope of each edge represents the value of $2\pi F_0$ in some interval between D8 branes, 
and the change of slope at each vertex equals $m_a$, the number of D8 branes in each group.  

If we compare the final configuration of branes with the initial configuration, with two groups of D8 branes well to the left or well to the right of the NS5 branes, we can establish the relation between the $(n_a,m_a)$ data and the partitions $\rho_{L,R}$ of $k$. As we carry a D8 brane from the $a$-th interval to the left, we create $a$ D6 branes ending on it. Similarly, if we bring it to the right, we create $N-a$ D6 branes. It is easy to show we can always separate the D8 branes in two groups, to be brought to the left or to the right in such a way that the NS5 branes live at $F_0=0$. 
 
These brane manipulations allow one to map every linear quiver of $N-1$ nodes with $N_f = 2 N_c$ at each node to a configuration of some $k$ 
D6 branes crossing $N$ NS5 branes, ending on two stacks $\rho_{L,R}$ of D8 branes. These original quivers have interesting Higgs branches, and 
we expect each of them to correspond to 6d SCFT with the labels $(N,k,\rho_L, \rho_R)$.
 
Conversely, the brane rearrangement we considered in this section can also be used to show that there is a lower bound on the number $N$ of NS5-branes compatible with 
the existence of a 6d SCFT with labels $(N,k,\rho_L, \rho_R)$. Consider moving away from the central stack as many NS5's as are needed to achieve a configuration where 
no D6-branes are ending on any D8's, as in the lower-right picture in figure \ref{fig:branes}. If there are not enough NS5-branes to achieve this,
we will be left with one or more D6 brane segments suspended between D8 branes. 

We expect that, as in lower dimensional examples, 
these D6 brane segments correspond to non-singular directions of the Higgs branch for the system. 
This could be shown, for example, by identifying the Higgs branch of the brane system with the space of solutions of the Nahm equations corresponding to supersymmetric configurations of the D6 brane world-volume theory. 

The theory $T^{A_N}_{A_k,\rho_L,\rho_R}$ which sits at the bottom of the flow 
initiated by the $\rho_{L,R}$ Higgs branch vevs, as in the previous section,
should not be an independent 6d SCFT in this case, but rather the product of some free hypermultiplets and a simpler 6d SCFT, 
obtained by removing the excess D6 branes from the system. 

We will see in section \ref{sec:expl} that the bound on the number of NS5-branes has a precise gravity counterpart (section \ref{ssub:bound-H}); 
the limit case where all NS5 branes need to be moved away will correspond to purely massive solutions (section \ref{sub:d8m}).


\subsection{Some generalizations} 
\label{sub:gen}

One can add a variety of other half-BPS defects to the original M-theory setup or to the IIA setup in order to produce other
families of 6d SCFTs. All the modification available in IIA constructions can be carried over to our holographic dual descriptions. 

An important class of examples arise from M5 branes sitting at the locus of an ADE singularity $R^4/\Gamma$. The singularity itself supports a seven-dimensional 
$G_\Gamma$ gauge theory and the M5 branes behave as domain walls in the world-volume of the gauge theory. 
Thus the resulting 6d SCFTs are characterized by a $G_\Gamma \times G_\Gamma$ flavor symmetry. 
If $\Gamma= D_n$, that has a IIA description in terms of an O6 plane added on top of the D6 branes. 
On the tensor branch, these theories will be associated to ortho-symplectic quivers. 

A second possibility is to consider M5 branes at the locus of an ``end of the  world'' $E_8$ boundary. 
Because of the $E_8$ gauge symmetry at the boundary, the M5 brane world-volume theories
is expected to have an $E_8$ flavor symmetry. It is likely possible to combine the $E_8$ boundary and an orthogonal ADE singularity, to find 6d theories with $E_8 \times G_\Gamma$ flavor symmetry. In IIA, the $E_8$ wall should reduce to an O8 wall combined with $8$ D8 branes to give a perturbative ${\rm SO}(16)$ flavor symmetry.
In the gauge theory, that should correspond to a terminal ${\rm USp}$ node in a quiver.
In IIA there are other variants for the O8 construction, described in \cite{hanany-zaffaroni-6d}. Again, the O8 planes could be easily added to the holographic dual description given in this paper.



\section{General results about AdS$_7$ vacua} 
\label{sec:ads}

We will now consider AdS$_7$ vacua of type II theories. In this section we will give a quick review of the general results in \cite{afrt}, which reduced the problem to a system of ODEs, and found several explicit solutions. In section \ref{sec:expl} we will then give a semi-empirical classification of such explicit solutions to the ODEs, finding a match with the brane configurations discussed in section \ref{sub:branes}.

\subsection{Metric and fluxes} 
\label{sub:mf}

Without loss of generality, we can write the ten-dimensional metric as a warped product
\begin{equation}
	ds^2_{10}= e^{2A} ds^2_{{\rm AdS}_7} + ds^2_{M_3}\ .
\end{equation}
The smallest possible superalgebra for a CFT$_6$ has an SU(2) R-symmetry; the internal space $M_3$ will thus be a fibration of a round $S^2$ over a one-dimensional space:
\begin{equation}\label{eq:met-r}
	ds^2_{M_3} = dr^2 + \frac1{16}e^{2A}(1-x^2)ds^2_{S^2} \ , 
\end{equation}
where $x$ is a function of $r$ only; the same is true of $A$ of the dilaton $\phi$. Locally, it turns out that supersymmetry reduces to a system of ODEs for these three quantities: 
\begin{equation}\label{eq:oder}
\begin{split}
	&\del_r \phi = \frac14 \frac{e^{-A}}{\sqrt{1-x^2}} (12 x + (2x^2-5)F_0 e^{A+\phi}) \ ,\\
	&\del_r x = -\frac12 e^{-A}\sqrt{1-x^2} (4+x F_0 e^{A+\phi}) \ ,\\
	&\del_r A = \frac14 \frac{e^{-A}}{\sqrt{1-x^2}} (4x - F_0 e^{A+\phi})\ .
\end{split}
\end{equation}

Globally, a priori $r$ could describe an interval or an $S^1$. It turns out that periodically identifying $r$ is inconsistent with (\ref{eq:oder}); this leaves us with the first option, which means $r \in [r_{\rm NP}, r_{\rm SP}]$; the labels stand for ``North Pole'' and ``South Pole'' respectively. In order to avoid boundaries, we see from (\ref{eq:met-r}) that $(1-x^2)$ has to go to zero at both $r_\pm$. It turns out that the poles are regular points with the boundary conditions
\begin{equation}\label{eq:bc}
	\left\{x=1,\ e^{A+\phi}=\frac4{F_0}\right\} \ \ {\rm at}\ r=r_{\rm N} \ ,\qquad
	\left\{x=-1,\ e^{A+\phi}=-\frac4{F_0}\right\} \ \ {\rm at}\ r=r_{\rm S} \ .
\end{equation}
Topologically, we have $M_3 \cong S^3$.

The rest of the solution is also completely determined, once (\ref{eq:oder}) is solved. $F_0$ is locally a constant, which flux quantization requires to be
\begin{equation}
	F_0 = \frac1{2\pi} n_0 \ ,
\end{equation}
for $n_0 \in \zz$. $F_2$ is given by
\begin{equation}\label{eq:F2}
	F_2 = q \,\left(\frac x4 F_0 e^{A+ \phi}-1\right){\rm vol}_{S^2}\ 
\end{equation}
where $q$ is a function of $r$ defined by 
\begin{equation}\label{eq:q}
	q \equiv \frac14 \sqrt{1-x^2}e^{A-\phi}=e^{-\phi} {\rm radius}(S^2_r)\ .
\end{equation}
In a massive region, where $F_0\neq 0$, and away from sources, we can then also determine $B$: since $dF_2 - H F_0 =0$, we have $H=d\frac{F_2}{F_0}$; in other words, 
\begin{equation}\label{eq:B}
	B= \frac{F_2}{F_0} + b \ ,
\end{equation}
where $b$ is a locally closed two-form. We should also recall flux quantization requires $F_0 = \frac1{2\pi} n_0$, $n_0 \in \zz$, and that the Page charge 
\begin{equation}\label{eq:n2}
	n_2 = \frac1{2\pi} \int_{S^2} (F_2-B F_0)  
\end{equation}
be integer too. We then see that 
\begin{equation}\label{eq:b}
	b=-\frac{n_2}{2 F_0}{\rm vol}_{S^2}\ .
\end{equation} 
In a region where $F_0=0$, we cannot use the expression (\ref{eq:B}); however, in such a region the solution is known explicitly, and one finds
\begin{equation}\label{eq:Bmassless}
	B_{F_0=0} = -\frac3{32}\frac{R^3}{n_2}\left( x - \frac{x^3}3\right)\ 
\end{equation}
where $R$ is a constant defined as $R^3 = -2 n_2 e^{2A}|_{x=0}$. (The massless solution has a lift to eleven dimensions, and $R$ is interpreted there as the radius of the internal $S^4$.)

We should also recall that, as usual, $B$ is a ``connection on a gerbe''. It transforms non-trivially on chart intersections. On $U\cap U'$, $B_U - B_{U'}$ can be a ``small'' gauge transformation $d \lambda$, where $\lambda$ is a 1-form. More generally, it can be a ``large'' gauge transformation, namely a two-form whose periods are integer multiples of $4\pi^2$. In our case, we can imagine to cover $S^3$ with several open sets of the form $U_i=(r_{i+},r_{i-})\times S^2$, where $S^2$ is the transverse sphere in (\ref{eq:met-r}). On an overlap we should then have $B_{U_i} - B_{U_{i-1}}= k_i\pi {\rm vol}_{S^2}$, which results in $\int_{S^3} H = B_{\rm N}- B_{\rm S}= (4\pi^2) \sum_i k_i$, in agreement with flux quantization for $H$. One way of taking care of this is to define the periodic variable  
\begin{equation} \label{eq:hatb}
	\hat b (r) \equiv \frac1{4\pi} \int_{S^2_r} B_2 
\end{equation}
which has period $\pi$. Large gauge transformations can then be thought of as values of $r$ where $b$ jumps: 
\begin{equation}\label{eq:bhatjump}
	\hat b \to \hat b + k \pi\ . 
\end{equation}
From (\ref{eq:b}) we then also see that 
\begin{equation}\label{eq:n2jump}
	n_2\to n_2 - k n_0\ .
\end{equation}


\subsection{Boundary conditions and branes} 
\label{sub:bc-branes}

As it turns out, there are no regular solutions without any D-brane sources. The only sources that can preserve R-symmetry are D6's, anti-D6's, or O6's, at the poles; and/or D8's\footnote{We will always allow the D8's to have a world-volume field-strength $f$ turned on; strictly speaking, these are then D8/D6 bound states. We will still refer to them as D8's, for simplicity.} wrapping an $S^2$ at a certain value $r_{\rm D8}$ of $r$. 

Having a D6, anti-D6 or O6 at the poles turns out to be exactly what happens when one follows the evolution of (\ref{eq:oder}) to $x=1$ or $-1$, but misses the value $e^{A+\phi}=\pm \frac4{F_0}$ from (\ref{eq:bc}). In other words, the asymptotic behavior for the system (\ref{eq:oder}) automatically gives the correct asymptotics for an D6, anti-D6 or O6 solution. 
If for example we consider $F_0>0$, one finds from (\ref{eq:oder}) that at $x=1$ (the North Pole) $A+\phi$ can go to either $\pm\infty$. For $A+\phi \to-\infty$ one finds 
\begin{equation}
	e^A \sim r^{1/3} \ ,\qquad e^\phi \sim r\ ,\qquad x \sim 1 + r^{4/3} \ ,
\end{equation}
which turns out to be the correct asymptotic behavior for an anti-D6 stack, in our coordinates. For $A+\phi\to \infty$, one gets
\begin{equation}
	e^A \sim r^{-1/5}\ ,\qquad e^\phi \sim r^{-3/5} \ ,\qquad x \sim 1 - r^{4/5}
\end{equation}
which is correct for an O6. Still for $F_0>0$, at $x=-1$ (the South Pole) $A+\phi$ can only go to $-\infty$, which correspond to a D6 stack. Although the correct asymptotic behavior is automatically achieved, one still has to impose that the number of branes in the stack is quantized. This can be done by recalling (\ref{eq:F2}); we simply have to impose that the function in front of ${\rm vol}_{S^2}$ is a half-integer. In particular, for the case $A+\phi \to -\infty$, the first term in (\ref{eq:F2}) vanishes and we see that we should impose
\begin{equation}\label{eq:qd6}
	q|_{x=\pm 1} = - \frac k2 \ ,
\end{equation}
where $k\in \zz$ is the number of (anti)D6 in the stack.

For the D8's one should be more careful: their position is fixed by supersymmetry.  Calling $(n_0,n_2)$ the values of these charges on one side of the D8 and $(n_0',n_2')$ the values on the other side, one finds\footnote{This can be derived in three ways: from continuity of $B$, from the Bianchi identity for $F_2$ in presence of D8 sources, or from a brane probe computation. All three give the same result; for more details, see \cite[Sec.~4.8]{afrt}.} the condition 
\begin{equation}\label{eq:qd8old}
	q|_{r=r_{\mathrm{D}8}}= \frac12 \frac{n_2' n_0 - n_2 n_0'}{n_0'-n_0}\ .
\end{equation}
It is perhaps clearer to rewrite (\ref{eq:qd8old}) in terms of data of the D8. The jumps in RR fluxes are related to the brane charges by $\Delta n_0 = n_{\rm D8}$ and $\Delta n_2 = n_{\rm D6}$. Let us define the `slope' of a stack by
\begin{equation}\label{eq:slope}
	\mu \equiv \frac{n_{\rm D6}}{n_{\rm D8}}\ .
\end{equation}
This is an integer, because the stack is made of many D8's all with the same D6 charge; intuitively, if this were not the case, the D8's with different charge would be stabilized at different values of $r$.
(\ref{eq:qd8old}) becomes
\begin{equation}\label{eq:qd8}
	q|_{r=r_{\mathrm{D}8}}= \frac12 (-n_2 + \mu n_0) = \frac12 (-n_2'+\mu n_0') \ .
\end{equation}
To see what this means, suppose for example that we are considering a D8 stack near the North Pole, in absence of D6-branes, so that we can take $n_2=0$. Then, recalling (\ref{eq:q}), (\ref{eq:qd8}) says that the radius of the D8 stack divided by the string coupling $g_s$ goes linearly with $\mu$, which is the D6 charge of one of the D8's in the stack, or in other words the number of D6's bound to the D8: $r_{D8}\sim g_s n_{D6}$. This is rather intuitive: increasing the number of D6's increases their mutual electrostatic repulsion, and makes the D8 puff up a little more. This is similar to what found in \cite{myers} for D2-brane probes (with D0 charge) in a constant $F_4$ background; in that context, too, the radius turns out to be linear in the number of D0's stuck to the D2's. Famously, in \cite{myers} this is interpreted with a non-abelian ``fuzzy sphere'' vacuum in the action for a D0 stack, again in an $F_4$ background. Such an analysis will work in a similar way in our context: upon T-dualizing six times, it turns into the statement that a stack of D6's can puff up into a D8 shell, in presence of an $F_{10}=* F_0$ RR flux. (The details of the computation would be changed by the fact that we are working in curved space and in presence of other fluxes as well.) Thus we learn that (\ref{eq:qd8}) can be interpreted as a Myers effect for a D6 stack.

\bigskip

In \cite{afrt}, the classification of AdS$_7$ solution was reduced to the system (\ref{eq:oder}), with boundary conditions (\ref{eq:bc}), and possible D8 jumps at positions fixed by (\ref{eq:qd8}). Several explicit solutions were found.

First of all, when no D8's are present, it is possible to have solutions with an arbitrary number $n_{\rm D6}$ of D6's at the SP and $n_{\rm \overline{D6}}$ of anti-D6's at the NP; these solutions have Romans mass $F_0$ and NS three-form $H$, restricted by the Bianchi identity to satisfy $n_{\rm D6}- n_{\rm \overline{D6}}= F_0 H$.\footnote{It is also possible to have O6's, in which case this constraint gets modified in the obvious way.} The particular case $F_0=0$ is nothing but a Hopf reduction of AdS$_7\times S^4/\zz_k$.

When one also considers D8-branes, the situation gets considerably more interesting, as we now discuss.



\section{Explicit solutions} 
\label{sec:expl}

We will now give a classification of explicit solutions to the system presented in section \ref{sec:ads}. As we will see, the most general solution is classified by the data of the D-branes present in the solution, and by the flux integer $N\equiv -\frac1{4\pi^2} \int H$, subject to the conditions described in subsections \ref{ssub:order}, \ref{ssub:bound-H}.

We will start in section \ref{sub:d8} by considering solutions with D8-branes such that there is a region where $F_0=0$. Solutions without such a region, which we will consider in section \ref{sub:d8m}, can be thought of as limits of those in section \ref{sub:d8}. In section \ref{sub:d8d6} we will then consider the case where also D6-branes are present at the poles.

\subsection{Solutions with D8's and a massless region} 
\label{sub:d8}

The boundary condition (\ref{eq:bc}) shows that, in order to obtain regular solutions, we should have $F_0>0$ at the North Pole and $F_0<0$ at the South Pole. In this section, we will assume that there is also a middle region where $F_0=0$. This is not necessarily true: one could jump from $F_0>0$ to $F_0<0$ directly through an appropriate stack of D8's. That case will be considered in section \ref{sub:d8m}. However, in a sense that case can be thought of as the limit where the massless region is shrinking to zero.

\begin{figure}[ht]
\centering	
	\subfigure[\label{fig:4d8as-a}]{\includegraphics[scale=.8]{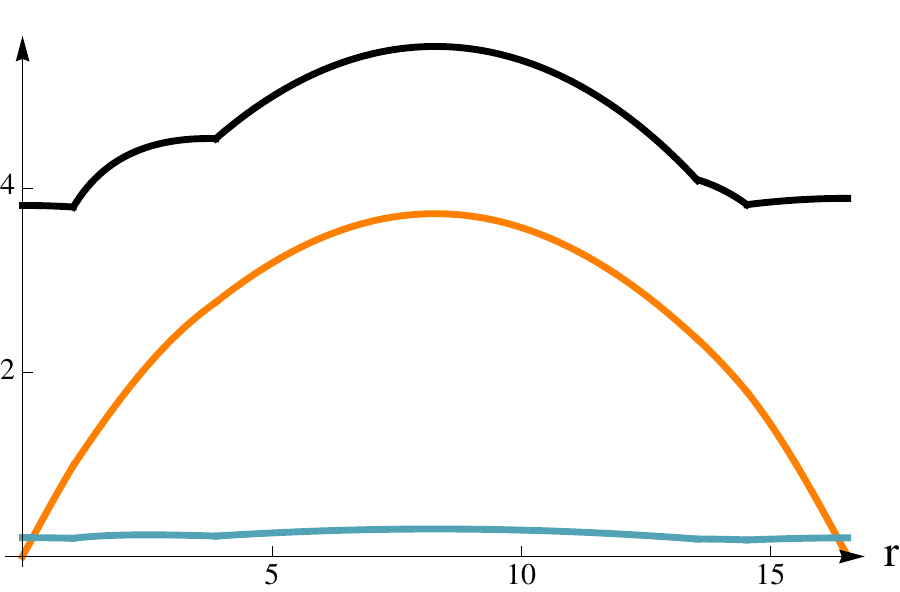}}
	\hspace{1cm}
	\subfigure[\label{fig:4d8as-b}]{\includegraphics[scale=.5]{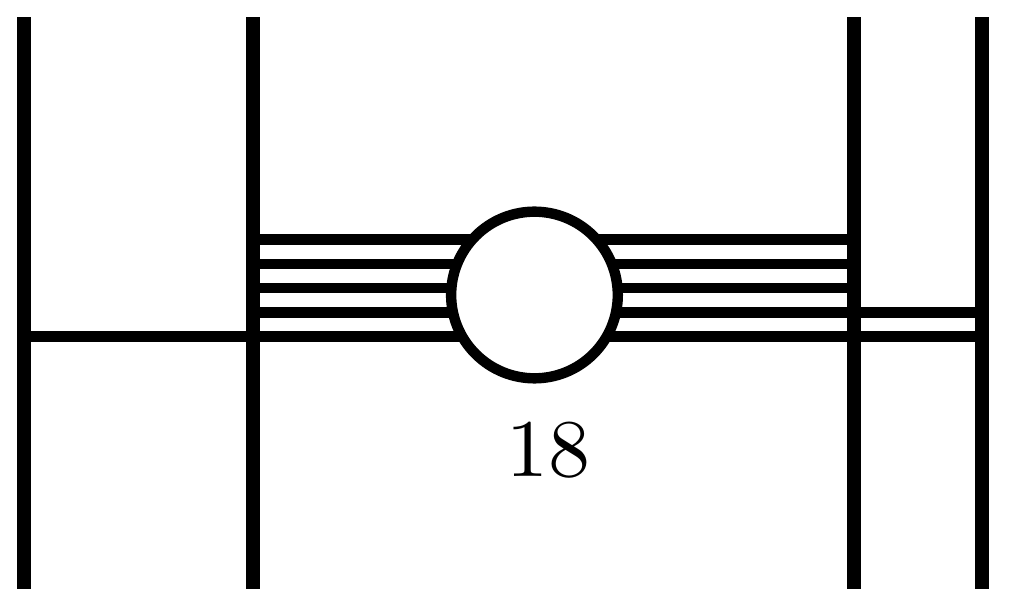}}
	\caption{Solution with four D8-brane stacks, placed asymmetrically around the equator. In \subref{fig:6d8s-a}, we plot the radius of the transverse $S^2$ (orange), the warping factor $e^{2A}$ (black; multiplied by $1/40$, also in all subsequent figures), and the string coupling $e^\phi$ (green).  The slopes are given, from left to right, by $\mu_i= \frac{n_{{\rm D6},i}}{n_{{\rm D8},i}}=\{1,4;-3,-2\}$. The central area around the equator has $F_0=0$, as described in the text. In \subref{fig:6d8s-b}, we see the corresponding brane configuration; the central circle represents a stack of 18 NS5-branes. (Here and in all the figures that follows, the number of D8 branes in the gravity solution is actually 5 times the number of D8's in the brane pictures. We do this so that the dilaton is small; increasing $n_{\rm D8}$ would make it even smaller. See discussion around (\ref{eq:nF0}).)}
	\label{fig:4d8as}
\end{figure}

\begin{figure}[h]
\centering	
	\subfigure[\label{fig:6d8s-a}]{\includegraphics[scale=.8]{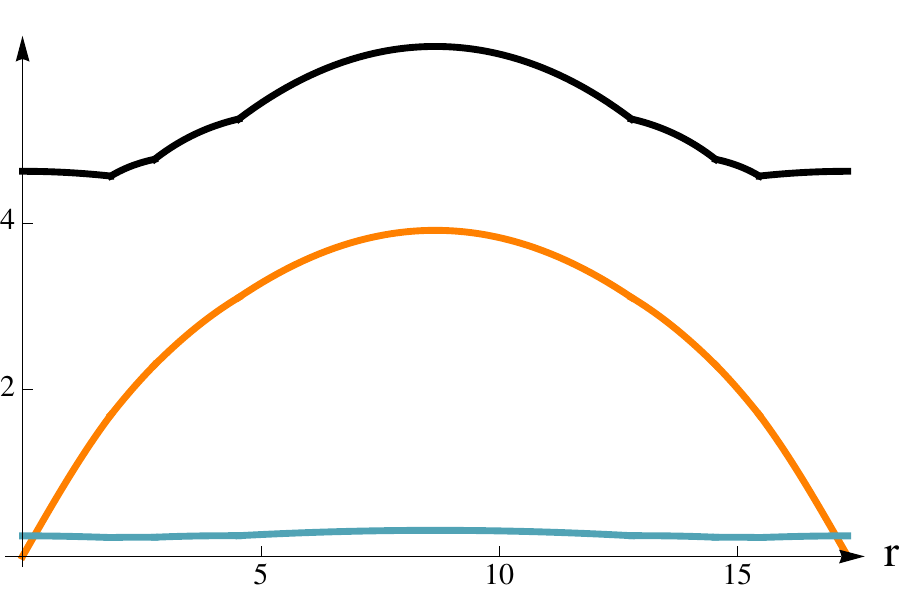}}
	\hspace{.4cm}
	\subfigure[\label{fig:6d8s-b}]{\includegraphics[scale=.5]{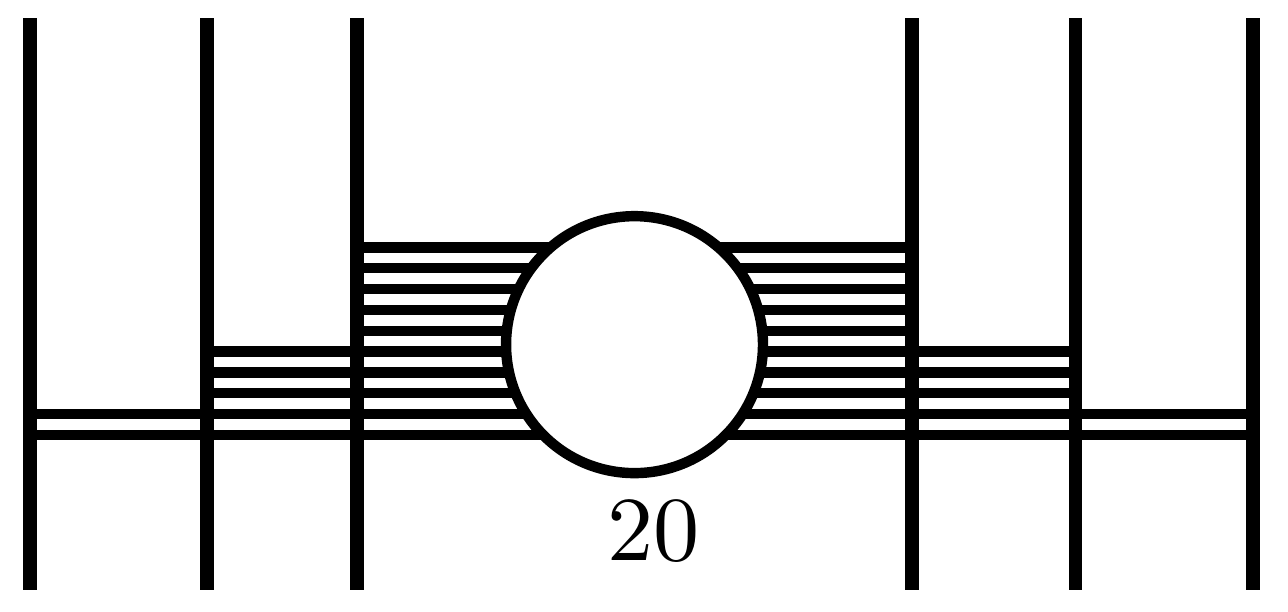}}
	\caption{Solution with six D8-brane stacks placed symmetrically around the equator.  The slopes are $\mu_i= \frac{n_{{\rm D6},i}}{n_{{\rm D8},i}}=\{2,3,5;-5,-3,-2\}$. The central area around the equator again has $F_0=0$. In \subref{fig:6d8s-b}, we see the corresponding brane configuration. The central circle represents a stack of 20 NS5-branes.}
	\label{fig:6d8s}
\end{figure}

One solution with a massless region, and with two D8 stacks, was obtained already in \cite[Sec.~5.3]{afrt}. In figures \ref{fig:4d8as-a} and \ref{fig:6d8s-a} we show examples with four and six stacks.

To find such numerical solutions, it is convenient to start at the poles, and work one's way towards the center. At each pole one needs to impose the boundary conditions (\ref{eq:bc}), and then follow the first-order system (\ref{eq:oder}). Notice that the boundary conditions (\ref{eq:bc}) have two degrees of freedom: they do not fix the values of $A(r_{\rm N})$ and $A(r_{\rm S})$. Starting for example from the North Pole, the evolution stops when the first D8 stack is met; namely, where the condition (\ref{eq:qd8}) is satisfied. From that point on, $F_0$ changes, and with it the system (\ref{eq:oder}). This is one of the angular points in figures \ref{fig:4d8as-a} and \ref{fig:6d8s-a}. 

Continuing this process, we are eventually going to reach the D8 stack just before the massless region. Now $q$ cannot change any more, because from (\ref{eq:oder}) we derive
\begin{equation}\label{eq:q'}
	\del_r q = \frac14 F_0 e^A\ .
\end{equation}
From this point of view, the position of the next D8 is now undetermined.
 
To get a clearer picture, let us think of the parameter space $\{A, \phi, x\}$. The solution we have just constructed from the North Pole corresponds to a trajectory on this space; it starts from the locus $q=0$ and it arrives on a two-dimensional surface $q=q_{\rm central}$.
Starting from the South Pole we can get a second trajectory, arriving on the same surface. The points reached on the surface  $q=q_{\rm central}$ depend on the parameters $A(r_{\rm N})$ and $A(r_{\rm S})$. By varying these parameters, we get two one-dimensional loci of points $l_{\rm N}$, $l_{\rm S}$ on this surface that can be reached by the northern and southern solutions. One possibility (which will be relevant for section \ref{sub:d8m}) would be to join the two solutions by looking for the intersection of the two loci $l_{\rm N}$ and $l_{\rm S}$. However, if we have a massless region, there will also be an evolution inside $q=q_{\rm central}$ (since, again, $q$ is constant when $F_0=0$). 

We conclude that, if we have a central massless region, we have a one-parameter family of supergravity solutions. This parameter, however, gets fixed when we impose flux quantization for $H$ --- namely, that $-\frac1{4\pi^2}\int H =N \in \zz$.

Implementing numerically this procedure leads to solutions such as those in figures \ref{fig:4d8as-a} and \ref{fig:6d8s-a}.

The general solution is characterized by the data of: 
\begin{enumerate}
	\item The number and charges of branes in each D8 stack: $(n_{\rm D8,i}, n_{\rm D6,i})$. 
	\item The integral $-\frac1{4\pi^2}\int H \equiv N \in \zz$.
\end{enumerate}

These data should be made large enough so that the solution is actually in the supergravity regime: both the curvature and the string coupling $e^\phi$ should be small. To arrange this, it is useful to notice some symmetries of the system (\ref{eq:oder}) (with the boundary conditions (\ref{eq:bc}) and the junction conditions (\ref{eq:qd8})). For example, in (\ref{eq:oder}) we see that $F_0$ always appears multiplied by $e^{A+\phi}$. Let us then rescale
\begin{equation}\label{eq:nF0}
	F_0 \to n F_0 \ ,\qquad \phi \to \phi - \log(n) \ .
\end{equation}
Locally, this turns a solution into a new solution; it is also compatible with the boundary conditions (\ref{eq:bc}). If we look at the condition on the position (\ref{eq:qd8}), (\ref{eq:q}), we see that the position of the D8's are not changed if we also rescale all the $n_{2,i} \to n n_{2,i}$, but leave the $\mu_i$'s invariant. Thus, once we find a solution, we can use (\ref{eq:nF0}) to generate new solutions where the dilaton is parametrically small. (In all the figures in this section and in section \ref{sub:d8m}, we have made use of this trick, in that the number of D8's and D6's in the solution is actually 5 times the one in the corresponding brane pictures.) 

Another symmetry we can use is 
\begin{equation}\label{eq:DeltaA}
	A\to A + \Delta A \ ,\qquad \phi\to \phi - \Delta A \ ,\qquad r \to e^{\Delta A} r \ ,\qquad x \to x \ .
\end{equation}
(Indeed, if one uses $A$ as a local coordinate, as in \cite[Eq.~(4.13)]{afrt}, one sees that the system is autonomous.) This leaves (\ref{eq:bc}) invariant; looking at (\ref{eq:qd8}), (\ref{eq:q}), we see that we should also rescale $n_{2,i}\to e^{2 \Delta A } n_{2,i}$ (and the $\mu_i$), while leaving the $n_{0,i}$ untouched. Using the expressions (\ref{eq:B}), (\ref{eq:Bmassless}), we see that the $B$ field also rescales in the same way, so that its flux integer $N \to e^{2 \Delta A} N$. We can then use (\ref{eq:DeltaA}) to generate new solutions where $A$ is arbitrarily large; from (\ref{eq:met-r}) we see that this makes the geometry larger, and hence the curvature arbitrarily small.

\subsubsection{Interpretation as near-horizon limit} 
\label{ssub:int}

It is natural to conjecture that the AdS$_7$ solutions correspond to the brane configurations we saw in section \ref{sub:branes}. The slope $\mu= \frac{n_{\rm D6}}{n_{\rm D8}}$ correspond to the number of D6's ending on a D8; the flux integer $N=-\frac1{4\pi^2}\int H$ should correspond somehow to the number of NS5-branes. For the solutions in figures \ref{fig:4d8as-b} and \ref{fig:6d8s-b}, we show the corresponding brane configurations in \ref{fig:4d8as-b} and \ref{fig:6d8s-b} respectively. 

\begin{figure}[ht]
	\centering
		\includegraphics[scale=.5]{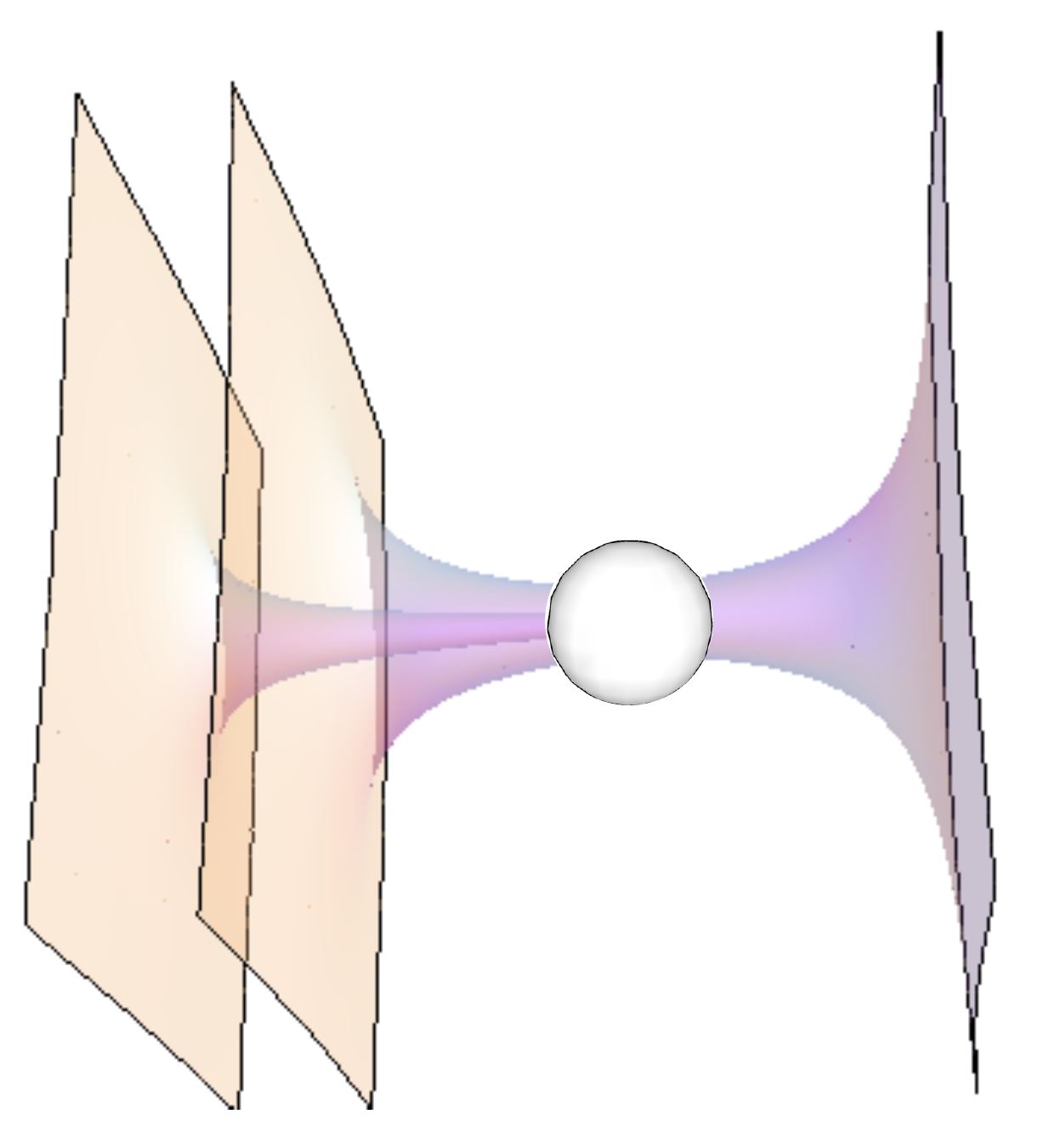}
	\caption{An artist's impression of a system with a few D8--D6 ``spikes''; the central sphere represents the NS5 stack. As one takes a near-horizon limit near the $N$ NS5's, they dissolve as usual leaving behind $N$ quanta of $H$ flux; the D8--D6 spikes should then become the D8--D6 bound states in the gravity solutions depicted in figures \ref{fig:4d8as-a}, \ref{fig:6d8s-a}. Intuitively, one can think of the intersections between the spikes and the central sphere as the location of the D8--D6 sources in the internal $S^3$ of the gravity solutions AdS$_7 \times S^3$.}
	\label{fig:spikes}
\end{figure}

The idea of this correspondence is that the AdS solutions come from an appropriate near-horizon limit of the brane configuration. We do not have a supergravity solution corresponding to the brane intersections,\footnote{For the NS5--D6 intersection, with no D8's, it is possible to obtain a solution valid near the D6's but otherwise at arbitrary distances from the NS5. This can be computed for example by starting from the M5 solution, and reducing it along the U(1) described in \cite[Sec.~5.1]{afrt}. It can also be derived directly in IIA using \cite{youm}, or using the formalism in \cite{janssen-meessen-ortin,imamura-D8}.} so we cannot really check this conjecture; however, it might work as follows. We can imagine putting the NS5's all on top of each other, and zooming in on the NS5--D6 intersection. One might imagine that this forgets about the D8's altogether. However, the classical pictures in figures \ref{fig:branes}, \ref{fig:4d8as-b}, \ref{fig:6d8s-b} (and those that will follow) are naive; the D6--D8 system is actually described by a Nahm pole as in (\ref{eq:nahm}), and should really be thought of as a single ``spike'' where the D6 gradually opens up into a D8. When performing the near-horizon limit, then, the NS5's get dissolved into $H$ flux; but the spikes are not scaled away, and remain as explicit sources even after the near-horizon limit --- see figure \ref{fig:spikes}. It is natural to think that these are indeed the D8 sources we see in our solutions, such as \ref{fig:4d8as-a}, \ref{fig:6d8s-a}, and those that we will see later.

The presence in the near horizon of the D8 brane sources at specific locations is somewhat similar to what one sees in the holographic duals of 3d ${\cal N}=4$ SCFTs in  \cite{assel-bachas-estes-gomis,aharony-berdichevsky-berkooz-shamir}.

As a further confirmation of this scenario, notice that the spikes are described in (\ref{eq:nahm}) as non-abelian solutions: the three matrices $t^i_\rho$ give an SU(2) representation, in general not irreducible; each of its irreducible sub-representations corresponds to a spike of different radius (for example, we see two of them on the left of figure \ref{fig:spikes}). In the near-horizon limit, these non-abelian spikes simply become fuzzy spheres; this agrees with our interpretation in section \ref{sub:bc-branes} of the D8--D6 bound states as the result of a Myers effect.

Since in the correspondence the location of the NS5's is unimportant, in figures \ref{fig:4d8as-b} and \ref{fig:6d8s-b} we have chosen to depict them all on top of each other, and to place them all in the massless region. As we will see more clearly in the fully-massive case in section \ref{sub:d8m}, actually the role of an NS5 is more or less played in the AdS solutions by the gauge transformations (\ref{eq:bhatjump}), (\ref{eq:n2jump}); placing those in the massless region is then convenient, since there they do not change $n_2$. 

\bigskip 

In our CFT discussion of section \ref{sec:cft}, we mentioned two constraints for a brane configuration: the ordering constraint (section \ref{sub:scft}), and the bound on the number of NS5-branes (section \ref{sub:branes}). We will now see that precisely these two constraints also appear for our AdS solutions.


\subsubsection{Ordering constraint} 
\label{ssub:order}

While looking for solutions, we have found that $F_0$ is always monotonous: in other words, there are only D8's, and no anti-D8's. Moreover, the D8's in the northern hemisphere have anti-D6's charge, while the ones on the souther hemisphere have D6 charge. As before, we label $n_{0,i}$ the value of $\frac{F_0}{2\pi}$ beyond the $i$-th D8 stack, and $n_{2,i}$ the value of (\ref{eq:n2}); then 
\begin{equation}\label{eq:incn0}
	n_{{\rm D8},i}=n_{0,i}-n_{0,i-1}<0 \ ,\qquad n_{{\rm D6},i}=n_{2,i}-n_{2,i-1}<0\ . 
\end{equation}

From this we can now also derive a constraint on the slopes $\mu_i = \frac{n_{{\rm D6},i}}{n_{{\rm D8},i}}$: from (\ref{eq:qd8}) we have
\begin{equation}\label{eq:qmu}
	2(q_{i+1}-q_i) = (-n_{2,i}+n_{0,i}\mu_i)-(-n_{2,i}+n_{0,i} \mu_{i-1})= n_{0,i}(\mu_i-\mu_{i-1})\ .
\end{equation}
Now, from (\ref{eq:q'}) we see that on the northern hemisphere $q^{i+1}> q^i$. From (\ref{eq:qmu}) then we see that
\begin{equation}\label{eq:incmu}
	\mu_i>\mu_{i-1}\ .
\end{equation}
Hence the D8 stacks have increasing slopes $\mu= \frac{n_{\rm D6}}{n_{\rm D8}}$. 

The ascending sequence stops in the central massless region, because there $n_0=0$, and $q$ remains constant (recall again (\ref{eq:q'})); hence there (\ref{eq:qmu}) is automatically satisfied, without yielding any constraints on the $\mu$'s before and after the massless region. In fact, vanishing of the total D6 charge implies that in the southern hemisphere the $\mu$'s are now negative; the same logic as above leads again to (\ref{eq:incmu}). In other words, the $\mu$'s are positive and growing in the northern hemisphere, and negative and growing in the southern hemisphere. This is indeed what happens in the explicit solutions of figures \ref{fig:4d8as} and \ref{fig:6d8s}.
	
The result (\ref{eq:incmu})	can be understood intuitively in a couple of ways. First we can appeal to our Myers effect interpretation described under (\ref{eq:qd8}) (and related in section \ref{ssub:int} to the non-abelian character of the Nahm poles (\ref{eq:nahm})). As we saw there, the radius (over the string coupling) $q = e^{-\phi} r_{S^2}$ of a D8--D6 bound state increases with the slope $\mu$. Moreover, (\ref{eq:q'}) shows that $q$ always increases when $F_0>0$. So, even though the geometry is non-trivially curved, we can always conclude that D8--D6 bound states with a higher D6 charge will be farther from the North Pole.
	
Another way of understanding (\ref{eq:incmu}) is this. Let us first think about the case with a single pair of D8 stacks, one in the northern hemisphere and one in the southern hemisphere. In this situation, the stacks would tend to ``slip'' towards the poles because of their tension; they are prevented from doing so from their mutual electric attraction due to their D6 charge (mediated by the $F_2$ field-strength). If we now go back to the case with many D8's with different slopes $\mu_i = \frac{n_{{\rm D6},i}}{n_{{\rm D8},i}}$, it is intuitively sensible that the D8's with larger D6 charge have a larger attraction towards their counterpart on the opposite hemisphere, and hence they should tend to be stabilized closer to the equator. This is what we found in (\ref{eq:incmu}).

Finally, and perhaps most importantly, (\ref{eq:incmu}) can be interpreted as the gravity dual of the ``ordering constraint'' reviewed in section \ref{sub:branes} for the D8's in the brane configurations. It was already observed in \cite[Sec.~2.2.2]{gaiotto-witten-3} that this could also be interpreted as an ordering constraint for the spikes of figure \ref{fig:spikes}. We now proved this constraint in the gravity duals, and we gave two simple physical interpretations for it.


\subsubsection{A bound on $\int H$} 
\label{ssub:bound-H}

The integer $N \equiv - \frac1{4\pi^2} \int H$ corresponds to a central massless region of different size. Empirically one sees that the size grows with $N$. This suggests that one should look at what happens in the two limits where the massless region shrinks to zero, and where it grows to fill everything. 

First we derive an expression for $\int H$, by putting together what we saw in section \ref{sec:ads}. We will assume to simplify our discussion that the ``equator'' $x=0$ falls into the massless region. Let us then first evaluate the integral $\int_{\rm north} H$ over the northern hemisphere, defined as the region from the North Pole up to equator. This is the sum over the northern massive region (from the north pole to the D8 stack, say D8$_n$, closest to the equator $x=0$) and over the northern massless region (from D8$_n$ to the equator $x=0$). Both can be evaluated by Stokes theorem, since we have expressions for $B$ in both regions: (\ref{eq:B}) and (\ref{eq:Bmassless}). Notice that we need not evaluate the integral separately in the stretch between each pair of D8 stacks. The expression (\ref{eq:B}) is valid up until the beginning of the massless region (namely at D8$_n$), including the integration constant (\ref{eq:b}). Thus we get
\begin{equation}\label{eq:Hnorth}
	\begin{split}
		\int_{\rm north} H &= \int_{r_{\rm N}}^{{\rm D8}_n} H + \int_{{\rm D8}_n}^{x=0} H \\
		 &= 4\pi\left[ q \left(\frac x4 e^{A+\phi}-\frac1{F_{0,n-1}}\right)-\frac{n_{2,n-1}}{2F_{0,n-1}}+\frac3{32}\frac{R^3}{n_{2,n}}\left( x - \frac{x^3}3\right)\right]_{{\rm D8}_n}\\
		&=4 \pi \left[-\pi \mu_n - \frac18 n_{2,n} x e^{A+\phi}+\frac3{32}\frac{R^3}{n_{2,n}}\left( x - \frac{x^3}3\right)\right]_{{\rm D8}_n}
	\end{split}
\end{equation}
A similar expression can be obtained for the southern hemisphere.

\begin{figure}[ht]
	\centering
		\includegraphics[width=0.5\textwidth]{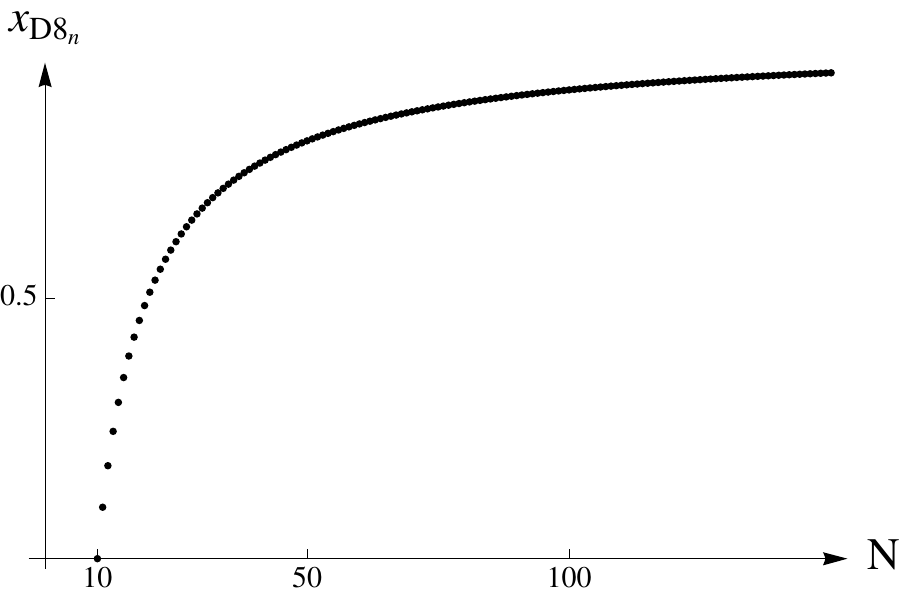}
	\caption{A plot of the position in $x$ of the last brane stack D8$_n$ versus the flux integer $N$, for the solution in figure \ref{fig:6d8s}.}
	\label{fig:N-x}
\end{figure}

We first examine the possibility that the massless region shrinks to nothing, which will be considered more thoroughly in section \ref{sub:d8m}. In this case, the last northern stack D8$_n$ is at $x=0$, and it coincides with the first stack D8$_{n+1}$ from the southern hemisphere. Setting $x=0$ in the contribution (\ref{eq:Hnorth}), we see that only the term with $\mu_n$ survives. Adding the contribution from the southern hemisphere, we get 
\begin{equation}
	\int H = -4\pi^2(\mu_n + \mu_{n+1})\ .
\end{equation}
Let us now suppose that the massless region exists but is very small.
We then have that $x_{{\rm D8}_n}\equiv \delta x \ll 1$. Recalling that $q_{{\rm D8}_n}=-\frac12 n_{2,n}$, after some manipulations we can approximate 
\begin{equation}\label{eq:intHappr}
	\int_{\rm north} H \sim - 4 \pi^2 \mu_n -\frac18 \pi \delta x e^{2A}|_{{\rm D8}_n}\ . 
\end{equation}
This shows that $-\int_{\rm north}H$ starts as $4\pi^2 \mu_n$, and then increases. Numerically one checks (see figure \ref{fig:N-x}) that the integral keeps growing as one makes the massless region larger, even beyond the approximation (\ref{eq:intHappr}). 

Putting this together with a similar discussion for the southern hemisphere, we have found that 
\begin{equation}\label{eq:Hbound}
	N = -\frac1{4\pi^2} \int H \ge \mu_n + \mu_{n+1}\ ,
\end{equation}
where $\mu_n$ and $\mu_{n+1}$ are the values of $\frac{n_{\rm D6}}{n_{\rm D8}}$ for the two D8 stacks (the $n$-th and $n+1$-th) surrounding the central massless region. 

There is no corresponding upper bound; $N$ can grow indefinitely, corresponding to a massless region which grows bigger and bigger, asymptotically eating up all the space, as shown in figure \ref{fig:N-x}.

In section \ref{ssub:int} we have identified $N$ as the gravity dual to the number of NS5-branes in the brane configurations of section \ref{sec:cft}. The bound we have found in this section is then the gravity dual of the bound we mentioned at the end of section \ref{sub:branes}. 
 

\subsection{Solutions with D8's without a massless region} 
\label{sub:d8m}

We have just derived a bound on $N$; for solutions that saturate that bound, the massless region shrinks to nothing. We will now study such cases in more detail.

\begin{figure}[ht]
\centering	
	\subfigure[\label{fig:1d8-a}]{\includegraphics[scale=.8]{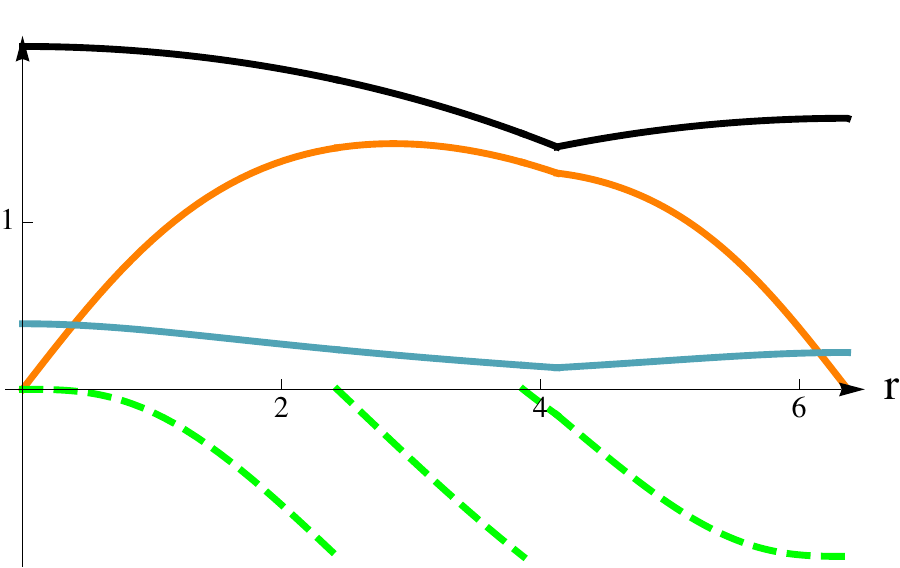}}
	\hspace{2cm}
	\subfigure[\label{fig:1d8-b}]{\includegraphics[scale=.6]{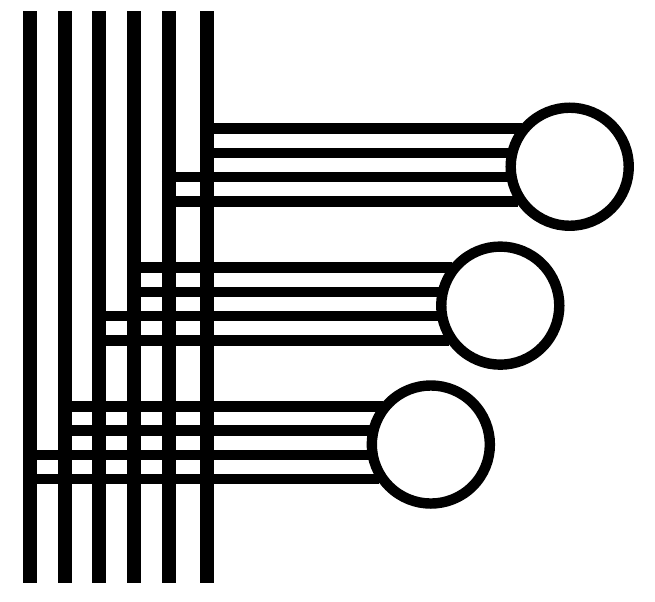}}
	\caption{Solution with a single D8-brane stack, with $\mu=\frac{n_{\rm D6}}{n_{\rm D8}}=2$. In \subref{fig:1d8-a}, along with the usual parameters (see figure \ref{fig:4d8as}), we show in dashed green the periodic variable $\hat b$ defined in (\ref{eq:hatb}). In \subref{fig:1d8-b}, we see the corresponding brane configuration (where each D8 and D6 should be imagined multiplied by $5$). There are 3 NS5's, corresponding to the three windings of $\hat b$ (which can be thought of as large gauge transformations) in figure \subref{fig:1d8-a}. Two D6 end on each D8, corresponding to $\mu=2$. Four D6's end on each NS5, corresponding to $n_0'=-4$ in the gravity solution.}
	\label{fig:1d8}
\end{figure}

The simplest such solution consists of one D8 stack. One was considered in \cite[Fig.4]{afrt}; we reproduce it in figure \ref{fig:1d8-a}. 

Let us review its construction here. Producing the supergravity solution itself is not unlike the strategy we described in section \ref{sub:d8}: we impose the appropriate regularity conditions (\ref{eq:bc}) at the North and South Poles, and we evolve towards the center. From both sides, we stop the evolution when we reach the appropriate value of $q=q_{\rm central}$ (as determined by (\ref{eq:qd8})). In general the solution obtained in this way will not be continuous. However, we can still vary the parameters $A(r_{\rm N})$, $A(r_{\rm S})$. Just as we described in section \ref{sub:d8}, by varying these we get two one-dimensional loci $l_{\rm N}$ and $l_{\rm S}$ in the space $\{A,\phi,x\}$. This time, however, we cannot connect them by a massless evolution; we will be forced to tune $A(r_{\rm N})$ and $A(r_{\rm S})$ so that $l_{\rm N}$ and $l_{\rm S}$ intersect. 

The upshot so far is that we have to spend both our free parameters in order for our solution to be continuous. We still have to impose flux quantization. It might seem that this should require tuning one additional parameter, which we do not have. Fortunately this is not the case. As we stressed in section \ref{ssub:bound-H}, the expression (\ref{eq:B}), (\ref{eq:b}) for the $B$ field is valid until we arrive at the massless region. If we never do, we can consider it valid everywhere. 

It is actually most convenient to work in a gauge where
\begin{equation}\label{eq:Bpoles}
	B\to 0
\end{equation}
at the poles; otherwise, by Stokes theorem, at the poles we get a delta-like singularity. This is not possible at both poles, unless we perform a large gauge transformation somewhere. These were actually also present for the massless solutions in section \ref{sub:d8}, but we could think of them as happening all within the massless region. When the solution is fully massive, this is not possible. We can for example perform the gauge transformations after the last D8 stack, before the South Pole. If the flux integers after the last D8 are $(n_{0,n},n_{2,n})$, performing (\ref{eq:n2jump}) will take them to $(n_{0,n},n_{2,n} - N n_{0,n})$. This will achieve (\ref{eq:Bpoles}) if and only if
\begin{equation}\label{eq:intgaugetr}
	n_{2,n} - N n_{0,n}=0 \ ;
\end{equation}
 indeed in that case from (\ref{eq:b}) we see that $b=0$, and the other term in (\ref{eq:B}) vanishes at $x=-1$. Thus, right before the last gauge transformation $n_{2,n}$ should be an integer multiple of $n_{0,n}$.

Actually, the role of these large gauge transformations is formally exactly the same as the role played by NS5-branes in the brane configurations of section \ref{sub:branes}. Indeed, one gauge transformation (\ref{eq:n2jump}) turns a flux integer $n_2$ into $n_2-n_0$; if, as in figures \ref{fig:4d8as-b}, \ref{fig:6d8s-b}, we associate the flux integer $n_2$ to the number of D6-branes, this is exactly like the behavior (\ref{eq:netD6}) for NS5-branes. For example, for the solution with one D8 in figure \ref{fig:1d8-a}, the brane configuration is shown in figure \ref{fig:1d8-b}.

\begin{figure}[ht]
\centering	
	\subfigure[\label{fig:2d8m-a}]{\includegraphics[scale=.8]{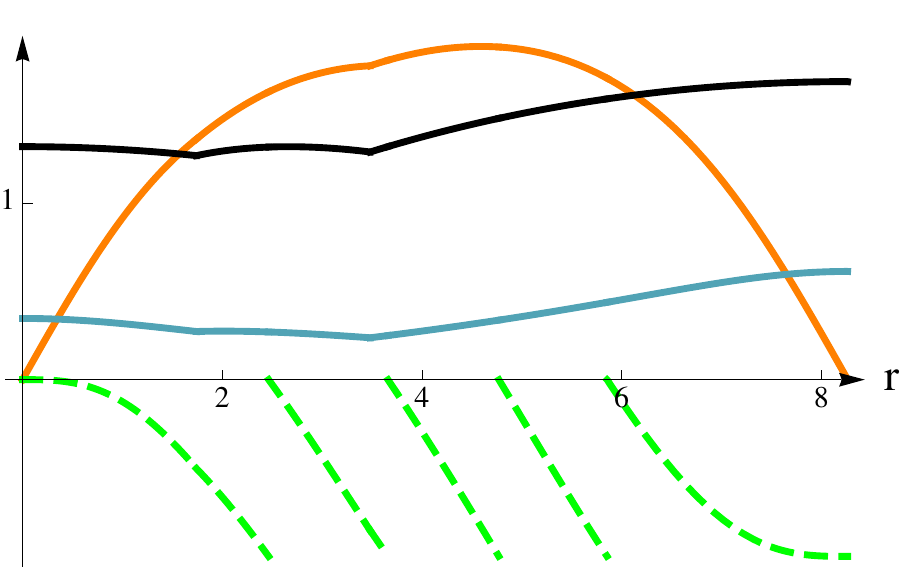}}
	\hspace{2cm}
	\subfigure[\label{fig:2d8m-b}]{\includegraphics[scale=.5]{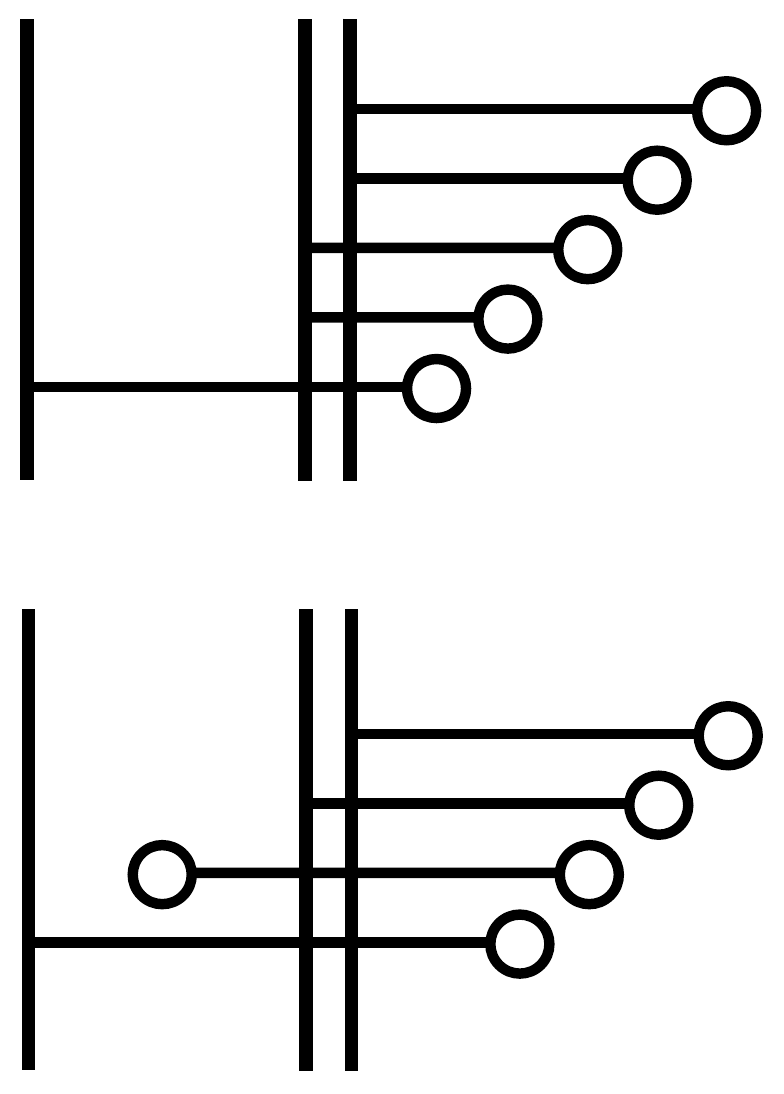}}
	\caption{Solution with two D8-brane stacks, and no massive region; once again we show in \subref{fig:1d8-a} the gravity solution; in \subref{fig:1d8-b} two possible brane configurations. On the gravity side, they correspond to different choices as to where to put the large gauge transformations.}
	\label{fig:2d8m}
\end{figure}

One can generalize the strategy we have described to find solutions with more than one stack, such as the one in figure \ref{fig:2d8m}. 

It is instructive to check why the quivers we obtain this way obey the s-rule. Let us do so, for example, for solutions with one stack, such as the one in figure \ref{fig:1d8}. Let us for simplicity call in this case $(n_0,n_2)$ the fluxes before the D8 stack, and $(n_0',n_2')$ afterwards. Let us also work in the gauge where $n_2=0$. Then $n_2'=n_{\rm D6}$, $n_0'-n_0=n_{\rm D8}$, $\mu = \frac{n_2'}{n_0'-n_0}$. From our general condition (\ref{eq:intgaugetr}), we see that $n_2'= N n_0'$. All this gives us \cite[Sec.~5.3]{afrt} $n_0= \left(1-\frac N\mu\right)n_0'$. Now, bearing in mind that $n_2\le 0$, $n_0 \ge 0$, $n_0'\le 0$, we have
\begin{equation}
	-n_2' \le - \mu n_{\rm D8}= \mu (n_0-n_0')\le -\mu n_0' \qquad \Rightarrow \qquad
	\mu \ge \frac{n_2'}{n_0'}= N \ ,
\end{equation}
which is indeed the usual s-rule for this case. A general argument along these lines can be given for the more general solutions with more than one D8 stack, such as the one in figure \ref{fig:2d8m}.


\subsection{Solutions with D8's and D6's} 
\label{sub:d8d6}

We finally present some solutions with D6's at one pole or the other. Such solutions can be found with a procedure similar to the one outlined in sections \ref{sub:d8} and \ref{sub:d8m}. The only additional subtlety is that we now cannot start the evolution from the pole on which the D6's are sitting; we have to start it somewhere else, and impose that (\ref{eq:qd6}) holds, by fine-tuning a free parameter in the solution. 

\begin{figure}[ht]
\centering	
	\subfigure[\label{fig:1d8d6-a}]{\includegraphics[scale=.8]{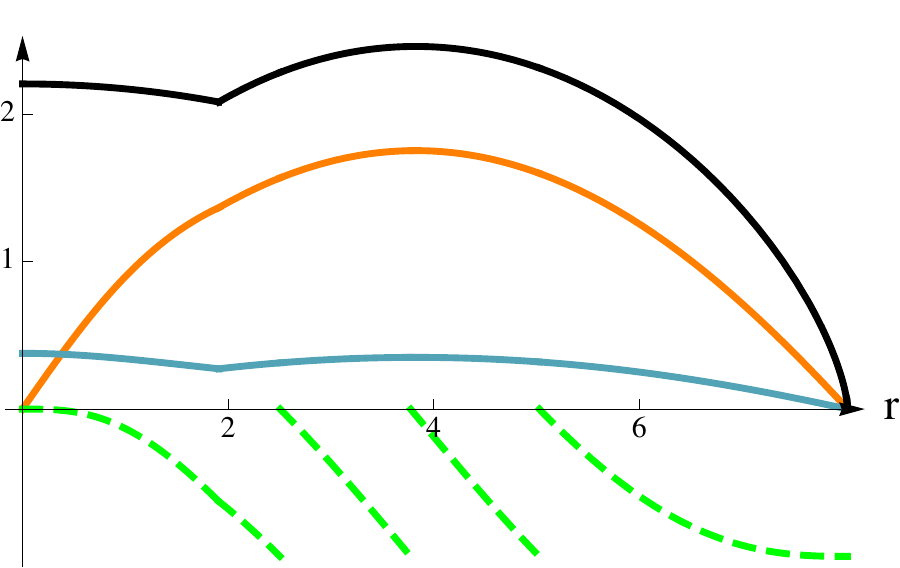}}
	\hspace{2cm}
	\subfigure[\label{fig:1d8d6-b}]{\includegraphics[scale=.8]{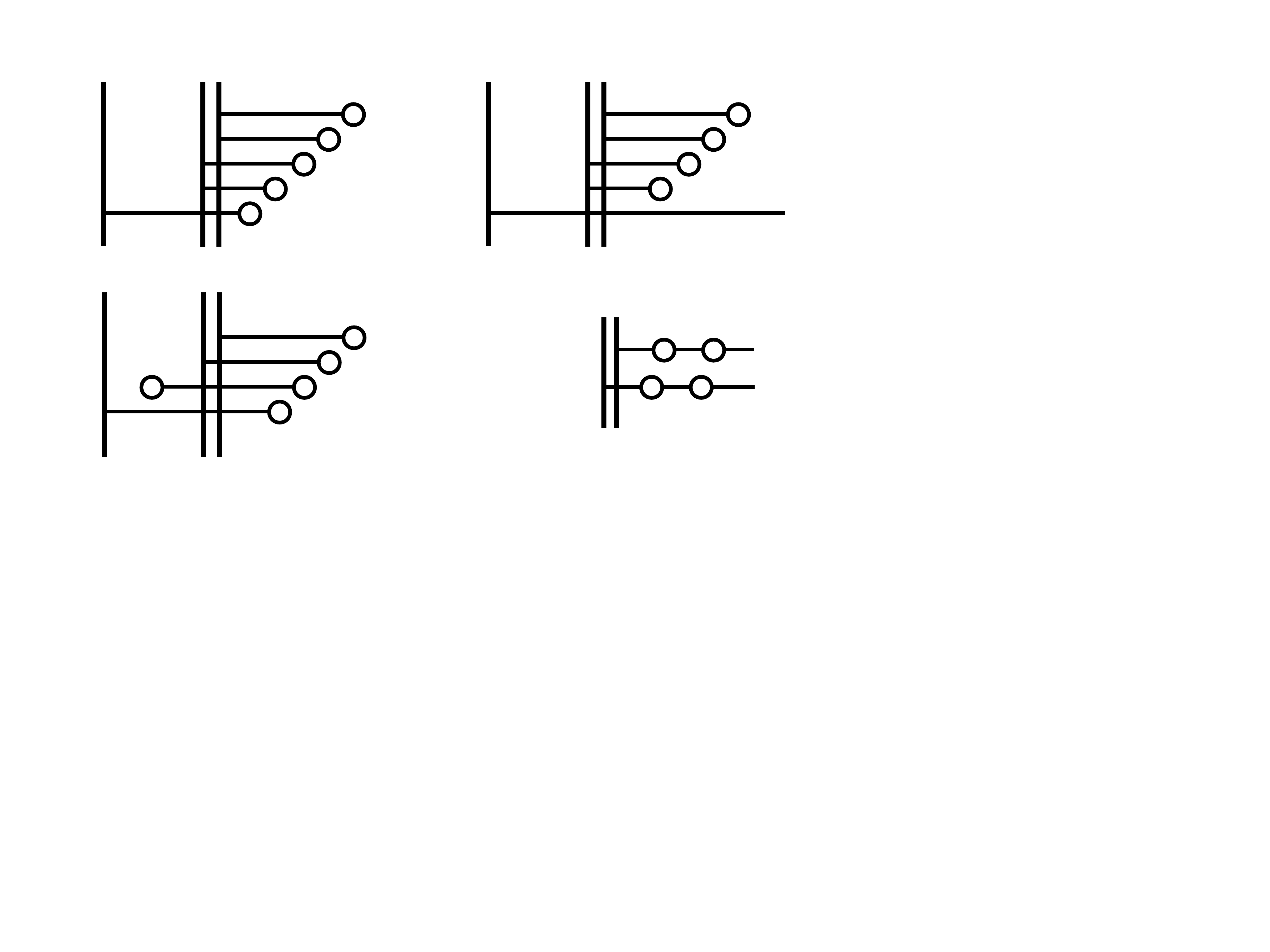}}
	\caption{A single D8 stack; on the left there is a massive region, on the right a massless region with D6-branes at the South Pole.}
	\label{fig:1d8d6}
\end{figure}

\begin{figure}[ht]
\centering	
	\subfigure[\label{fig:2d8md6-a}]{\includegraphics[scale=.8]{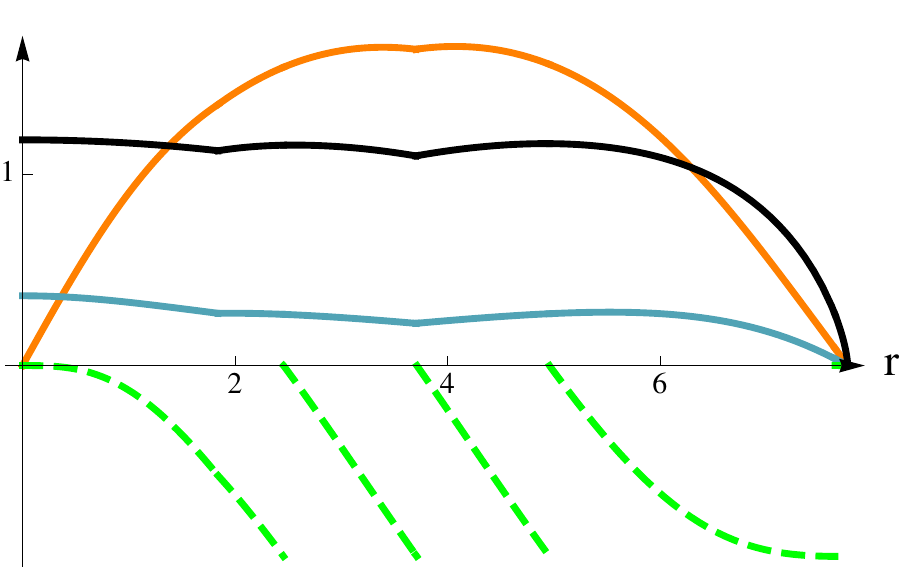}}
	\hspace{2cm}
	\subfigure[\label{fig:2d8md6-b}]{\includegraphics[scale=.6]{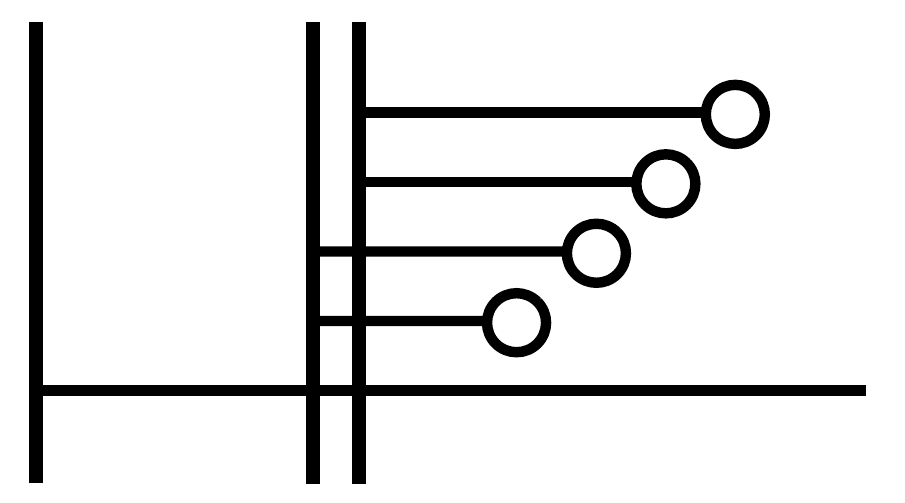}}
	\caption{This is a variation on figure \ref{fig:2d8m}; this time at the South Pole we place a D6.}
	\label{fig:2d8md6}
\end{figure}

We see two examples in figures \ref{fig:1d8d6} and \ref{fig:2d8md6}. In both cases, the D6's are at the South Pole $r=r_{\rm S}$. Their presence is signaled, as in \cite[Sec.~5.2]{afrt}, by the fact that both $e^A$ and $e^\phi$ go to zero, rather than going to a constant. As we mentioned in section \ref{sub:bc-branes}, the system (\ref{eq:oder}) is such that the correct asymptotics for a D6 singularity are automatically obtained.

It is finally also possible to have solutions with D6's and anti-D6's at both poles. 
Notice that we do not expect the D6 branes to represent something different than 
a stack of D8 branes, each of D6 brane charge $1$. Both will be dual to the same 6d SCFT.


\section{Degrees of freedom} 
\label{sec:free}

In this section, we are going to discuss the number of degrees of freedom for some of the CFT's we have been considering. 

\begin{figure}[ht]
\centering	
	\subfigure[\label{fig:free-no}]{\includegraphics[scale=.7]{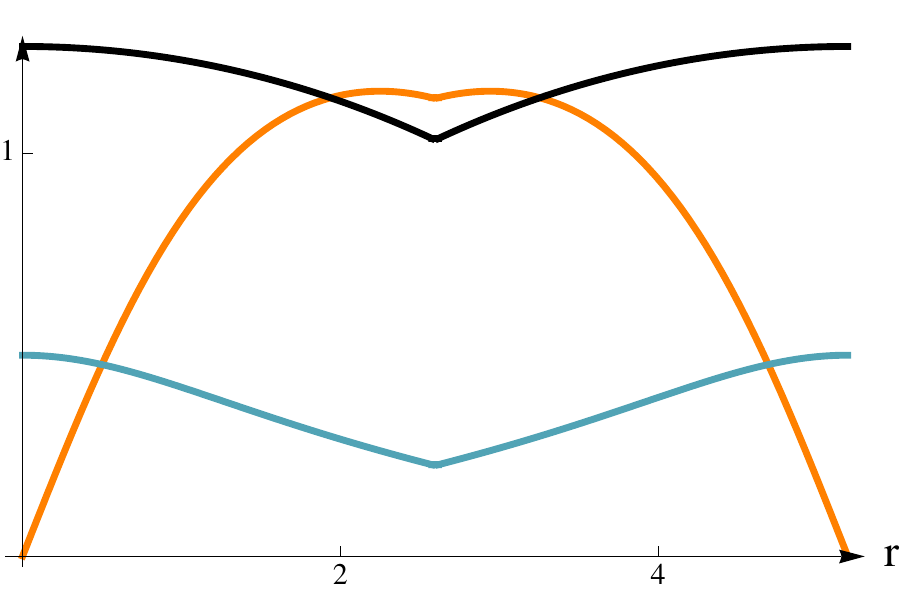}}
	\hspace{2cm}
	\subfigure[\label{fig:free-large}]{\includegraphics[scale=.7]{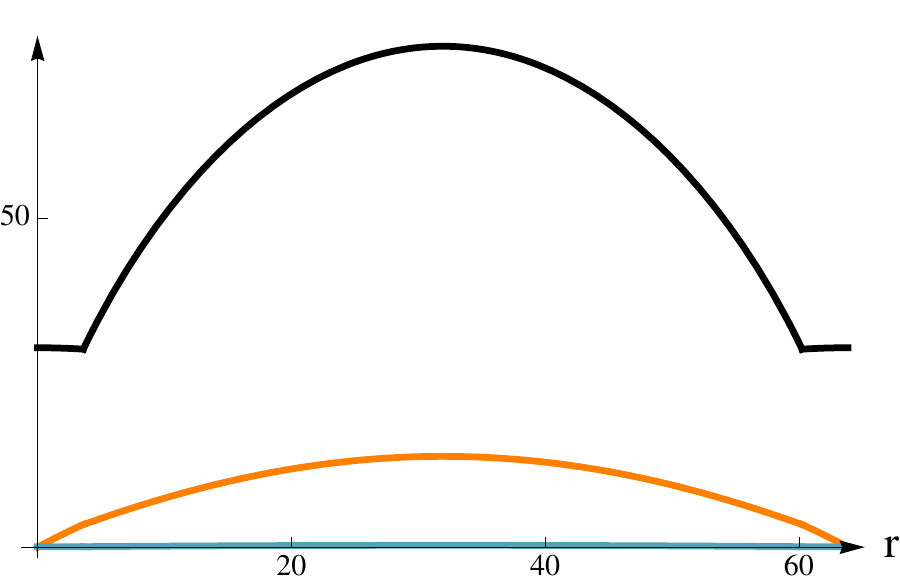}}
	\caption{The two limits we consider in this section; in \subref{fig:free-no} we see the case where the two D8 stacks coincide and the massless region has shrunk to nothing, while in \subref{fig:free-large} we see the case where the massless region is very large and takes up almost all the space.}
	\label{fig:free}
\end{figure}

In general, such an estimate can be obtained by putting the theory at a finite temperature, and computing its free energy. From the point of view of the holographic dual, this corresponds to computing the area of a sphere at a fixed radius in AdS:
\begin{equation}
	{\cal F}\sim\frac{\rm Area}{G_{\rm N}} \sim \frac{R^5 {\rm vol}(M_3)}{g_s^2}\ . 
\end{equation}
This reproduces the expected behavior $\beta {\cal F} = {\cal F}_0 {\rm Vol} T^{5}$; the coefficient ${\cal F}_0$ can then be taken to be as a measure of the number of degrees of freedom. In our case, however, neither $R$ nor $g_s$ are constant over the internal space. $R$ should get replaced by $e^A$, and $g_s$ by $e^\phi$; both should then be integrated over $M_3$:
\begin{equation}\label{eq:free}
	{\cal F}_0=\int e^{5A-2 \phi} {\rm vol}_3\ . 
\end{equation}
The AdS$_7$ vacua of section \ref{sec:ads} are known only numerically, and computing (\ref{eq:free}) explicitly is hard. We are going to consider a simple case: two symmetric stacks, with a massless region (see figure \ref{fig:free}). The solutions are characterized by three integers: $n_0$ and $n_2 \equiv k$ in the massive region around the North Pole, and $N= -\frac 1{4\pi^2} \int H $. In this situation, we are going to estimate ${\cal F}_0$ in two different limits: the case where the massless region is occupying most of the volume of $M_3$, and the opposite case where it shrinks to nothing; see figures \ref{fig:free-no} and \ref{fig:free-large}. As a warm-up, we are going first to consider the case where it takes up \emph{all} of the space: namely, the massless solution. 

\subsection{Free energy for the massless solution} 
\label{sub:free-massless}

As we mentioned earlier, the massless solution is actually nothing but the reduction of AdS$_7 \times S^4/\zz_k$. It would be easiest to compute the free energy directly in eleven dimensions; however, we compute it here in ten dimensions, as a warm-up to the later case where the massless region takes up most but not all of the space.
 
In this case, specializing \cite[Eq.~(4.18)]{afrt} to $F_0=0$, we learn that
\begin{equation}
	H = -6 e^{-A} {\rm vol}_3 \ .
\end{equation}
Moreover, from \cite[Eq.~(5.4)]{afrt} we see $e^{3A-\phi}=R^3$, a constant (that in eleven dimension is interpreted as the radius of $S^4$). Thus
\begin{equation}\label{eq:Fmassless}
	{\cal F}_0 = \int e^{5A-2 \phi} {\rm vol}_3 = -\frac16 R^6 \int H = \frac23 \pi^2 N R^6\ . 
\end{equation}
On the other hand, using (\ref{eq:Bmassless}), we can also compute easily $N=\frac{R^3}{4\pi k}$. This gives 
\begin{equation}\label{eq:N3k2}
	{\cal F}_0 = \frac{32}3 \pi^3 N^3 k^2\ .
\end{equation}

(\ref{eq:N3k2}) is the number of degrees of freedom of the $(1,0)$ $T^{A_N}_{A_{k}}$ theory. 
This should come about by considering some kind of orbifold of the $(2,0)$ theory.\footnote{We thank L.~Rastelli for an illuminating discussion about this.} 
If we think of the $(2,0)$ theory to be somehow described by ``cubic matrices'' of size $N$, in order to orbifold it we will have to start from cubic matrices of size $N k$, and then impose invariance under $\zz_k$, which will somehow divide the number of degrees of freedom by 
$k$; this results in $\frac1k (N k)^3  = N^3 k^2$, in agreement with (\ref{eq:N3k2}). A similar computation for an ordinary quiver (such as for the circular quiver describing 
D3-branes at the tip of a $\cc^2/\zz_k \times \cc$ singularity) 
correctly yields $\frac1k (Nk)^2= N^2 k$.


\subsection{No massless region} 
\label{sub:free-no}

We will now consider the case where there is no massless region, as in figure \ref{fig:free-no}. This is actually the case we already mentioned in section \ref{ssub:bound-H}, saturating the bound (\ref{eq:Hbound}): thus in this case $N=2\mu=2 \frac{k}{n_0}$.

Even though the massive solution is only known numerically, it depends on the flux parameters in a simple way, as we saw already in section \ref{sub:d8}. We can use the symmetries (\ref{eq:nF0}), (\ref{eq:DeltaA}); it is convenient to use $x$ as a local coordinate, since it is invariant. The solution that starts from the boundary condition (\ref{eq:bc}) can then be written as 
\begin{equation}\label{eq:solx}
	A= A_0 + \tilde A(x) \ ,\qquad \phi= -A_0+\log\left(\frac4{F_0}\right) + \tilde \phi(x)
	\ .
\end{equation}
For the solution with no massless region in figure \ref{fig:free-no}, the northern solution stops at the D8, which is located at $x=0$. Moreover, at that point (\ref{eq:qd8}) should hold; this tells us that
\begin{equation}\label{eq:A0n2F0}
	2k= e^{A-\phi}|_{D8}= \frac14 e^{2A_0}F_0 e^{\tilde A(x) - \tilde \phi(x)}|_{D8}\ . 
\end{equation}
This fixes the dependence of $A_0$ on the fluxes. Thus (\ref{eq:solx}) fixes the dependence of the whole solution on the flux integers. 

We can now turn to the free energy. Using (\ref{eq:met-r}), (\ref{eq:oder}), we can compute the volume form in terms of $dx$. This gives
\begin{equation}\label{eq:Fintmassive}
	{\cal F}_0 \sim \int e^{5A-2 \phi} {\rm vol}_3 = \frac14\int_0^1 dx \sqrt{1-x^2} \frac{e^{8A-2 \phi}}{4+x F_0 e^{A+ \phi}} = \frac{F_0^2}{64} e^{10 A_0}\int_0^1 dx \sqrt{1-x^2} \frac{e^{8\tilde A(x)-2 \tilde\phi(x)}}{4+x F_0 e^{A+ \phi}}\ .
\end{equation}
The integral is now a fixed constant; thus the dependence on the fluxes is captured by ${\cal F}_0\sim n_0^2 e^{10 A_0}$. Moreover, (\ref{eq:A0n2F0}) tells us $e^{2A_0}\sim \frac {k}{n_0}$. (The solution is then weakly curved when $k \gg n_0$ --- (\ref{eq:solx}) then also shows it is weakly coupled.) Altogether, we get
\begin{equation}\label{eq:Fmassive}
	{\cal F}_0\sim n_0^2 \left(\frac {k}{n_0}\right)^5 = \frac{k^5}{n_0^3}\ .
\end{equation}

We can compare this result to the quivers one gets upon a tensor-branch deformation, as in the horizontal arrows in figure \ref{fig:quivers}. As in those figures, we can move enough NS5's out of the central massless region so as to have no D6's end on any D8's. The case considered in this subsection corresponds to having no NS5's in the central region at the end of this process. The quiver one finds on the tensor branch has two tails which each consist of gauge groups of ranks $n_0$, $2 n_0$, \ldots $ \mu n_0=k$. The number of degrees of freedom of such a theory should go like $\sum_{j=1}^{\mu} (j n_0)^2 \sim n_0 \mu^3$. In the regime of validity of our solutions ($\mu= \frac{k}{n_0}\gg 1$) this is much less than the result ${\cal F}_0\sim n_0^2 \mu^5$ from (\ref{eq:Fmassive}). Such a large loss of degrees of freedom indicates that the much interesting dynamics is lost in the tensor-branch flow. 


\subsection{Large massless region} 
\label{sub:free-large}

We will now consider the case where the massless region takes up almost all of $M_3$, as in figure \ref{fig:free-large}. This will correspond to taking $N=-\frac1{4\pi^2}\int H $ very large, in a sense we will quantify shortly. 

Our starting assumption will be that $x$ at the northern D8 is very close to 1, which is the value it takes at the pole:
\begin{equation}
	x|_{D8}\sim 1 - \delta x \ . 
\end{equation}  
We can again use the ``universality'' of the regular massive solution we pointed out in (\ref{eq:solx}). This time we need to know a little more about the solution. Since it is only relevant from $1$ to $1-\delta x$, if we rewrite (\ref{eq:oder}) using $x$ as a variable we can study the solution perturbatively around $x=1$. Skipping the details of the analysis, we get: 
\begin{equation}\label{eq:pertx}
	A_{\rm D8} = A_0 - \frac1{24} \delta x - \frac1{32} \delta x^2 + \ldots \ ,\qquad
	\phi_{\rm D8} = -A_0 + \log\left(\frac4{F_0}\right) - \frac58 \delta x - \frac{13}{96}\delta x^2 + \ldots \ .
\end{equation}
We can now impose (\ref{eq:qd8}), and use it derive $e^{2A_0}$ perturbatively in $\delta x$:
\begin{equation}
	e^{2A_0}= -4\sqrt2 \frac{k}{F_0 \sqrt{\delta x}}\left(1-\frac13 \delta x -\frac1{72} \delta x^2+\ldots\right)\ .
\end{equation}
Before we evaluate the free energy, we should learn how to relate the flux integer $N$ to these parameters. This computation is quite similar to (\ref{eq:Hnorth}). In this case $n_{2,n-1}=0$ (since there is only one northern D8 stack). We get:
\begin{equation}\label{eq:Npert}
	-4\pi^2 N = \int H = -4\pi \left(\frac x4 e^{A+\phi} - \frac1{F_0} \right) + \int_{F_0=0} H = 8 \pi \left[\frac{k}{F_0}\left(\frac 56 \delta x -\frac{13}{30} \delta x^2 \right)+\frac 3{32}R^3{k}\left(\frac23 - \delta x^2\right)\right]\ .
\end{equation}
For the massless region we have used the fact that we know the solution explicitly, in terms of the integration constant $R^3= e^{3A-\phi}$ that we recalled in section \ref{sub:free-massless}. This can be determined in terms of our parameters by using another feature of the massless solution, namely that $R^3=-2k \frac{e^{2A}}{\sqrt{1-x^2}}$ (see \cite[Eq.~(5.6)]{afrt}). Expanding this perturbatively in $x$, and using (\ref{eq:Npert}), we finally determine $\delta x$:
\begin{equation}
	\delta x = -\frac{2k}{n_0 N}\left(1+ \frac13 \frac{k}{n_0 N}+\ldots\right)
\end{equation}
which shows that we are working in the regime where 
\begin{equation}
	N \gg \frac{k}{n_0}\ .
\end{equation}

We can finally evaluate the free energy. The contribution from the massless region is similar to the one we saw in (\ref{eq:Fmassless}), except that now the integral of $H$ is only performed from the northern D8 stack to the southern one:
\begin{equation}
	\int_{F_0=0} e^{5A-2 \phi} {\rm vol}_3 = -\frac16 R^6\int_{1-\delta}^{-1+\delta x} H = \frac{32}3 \pi^3 N^3 k^2 -\frac{256}3 \pi^3 N \frac{k^4}{n_0^2}+ \ldots \ .
\end{equation}
The contribution from the massive region can be obtained as in (\ref{eq:Fintmassive}), with now a very small integration interval:
\begin{equation}
\begin{split}
	\int_{F_0\neq 0} e^{5A-2 \phi} {\rm vol}_3 &= \frac14\int_1^{1-\delta x} dx \sqrt{1-x^2} \frac{e^{8A-2 \phi}}{4+x F_0 e^{A+ \phi}} \\
	&= \frac{\sqrt{2}}{3 \cdot 256} e^{10 A_0} F_0^2 \delta x^{3/2} + \ldots = -\frac{128}3 \pi^3 N \frac{k^4}{n_0^2}+ \ldots \ .
\end{split}
\end{equation}
Putting the two contributions together:
\begin{equation}\label{eq:Falmostmassless}
	{\cal F}_0 \sim \frac{32}3 \pi^3 N^3 k^2 - 128 \pi^3 N \frac{k^4}{n_0^2}+ \ldots\ .
\end{equation}

Thus the first correction to the free energy of the massless case (\ref{eq:N3k2}) is linear in $N$. One might think that this makes it compete with the quantum corrections of the massless case, which also go linearly in $N$; however, the coefficient $\frac{k^4}{n_0^2}$ dominates when for example $k \gg 1$. 

One can also again compare the result (\ref{eq:Falmostmassless}) with the number of degrees of freedom of a theory obtained by a tensor-branch deformation, as in section \ref{sub:free-no}. We can again move enough NS5's away from the central massless region so as to free all the D8's (so that no D6's end on them any more). The number of such NS5's is $2 \mu$; in section \ref{sub:free-no} this would leave no NS5's in the middle, but in the present case $N\gg \mu$, and the theory is expected to have $k^2(N-\mu)^3$ degrees of freedom (plus a contribution $n_0^2 \mu^3$ due to the two quiver ``tails'', similarly as in section \ref{sub:free-no}; except this is now much smaller than the leading term). Expanding this at large $N$, the first term $k^2 N^3$ is present in (\ref{eq:Falmostmassless}), but the next term $-k^2 N^2 \mu$ is absent. So (\ref{eq:Falmostmassless}) is again larger than the number of degrees of freedom of its tensor-branch deformation, as it should be.



\section*{Acknowledgments}
We would like to thank F.~Apruzzi, M.~Fazzi, D.~Rosa, L.~Rastelli, A.~Zaffaroni for interesting discussions. A.T.~is supported in part by INFN, by the MIUR-FIRB grant RBFR10QS5J ``String Theory and Fundamental Interactions'', and by the European Research Council under the European Union's Seventh Framework Program (FP/2007-2013) -- ERC Grant Agreement n. 307286 (XD-STRING).
The research of DG was supported by the Perimeter Institute for Theoretical Physics. Research at Perimeter Institute is supported by the Government of Canada through Industry Canada and by the Province of Ontario through the Ministry of Economic Development and Innovation.


\bibliography{at}
\bibliographystyle{at}

\end{document}